\documentclass[12pt]{article}
\usepackage{amsmath,epsfig,amssymb,eufrak}
\usepackage{amscd,latexsym}
\usepackage{color}
\usepackage{bbm}

\date{}

\def\lsi{\raise0.3ex\hbox{$<$\kern-0.75em\raise-1.1ex\hbox{$\sim$}}}
\def\gsi{\raise0.3ex\hbox{$>$\kern-0.75em\raise-1.1ex\hbox{$\sim$}}}

\newcommand{\R}{{R_\mathrm{det}}}

\input macros.sty      
\input eqalign.sty     
\newcommand{\bearr}{\begin{eqnarray}}
\newcommand{\eearr}{\end{eqnarray}}

\def\myrewritetimestamp Time-stamp: <#1 #2 #3>{#1 #2}

\def\zeroPadTwo#1{
  \ifnum #1<10 0\fi    
  #1
}
\newcount\minute        
\newcount\hour          
\newcount\hourMins  
\def\now{
  \minute=\time    
  \hour=\time \divide \hour by 60 
  \hourMins=\hour \multiply\hourMins by 60
  \advance\minute by -\hourMins 
  \zeroPadTwo{\the\hour}:\zeroPadTwo{\the\minute}%
}

\begin{document}

\begin{titlepage}

\title{
  {\vspace{-0cm} \normalsize
  \hfill \parbox{40mm}{DESY/05-195\\
                       HU-EP-05/45\\
                       October 2005}}\\[10mm]
Scaling test of fermion actions \\ in the Schwinger model}  

\author{N. Christian$^{\, 1}$,
K. Jansen$^{\, 1}$,
K. Nagai$^{\, 1}$
and B. Pollakowski$^{\, 1,\, 2}$
\vspace*{0.7cm} \\
$^{1}$   John von Neumann Institute for Computing,\\
Platanenallee 6, 15738 Zeuthen\\
\vspace{0.6cm}
\\
$^{2}$ Institut f\"{u}r Physik,
Humboldt Universit\"{a}t zu Berlin,\\
Newtonstr.\ 15, 12489 Berlin
}

\maketitle

\begin{abstract}
\vspace{0.75cm}
We discuss the scaling behaviour of different fermion
actions in dynamical simulations of the 2-dimensional 
massive Schwinger model. We have chosen
Wilson, hypercube, twisted mass and overlap fermion actions.
As physical observables, the pion mass
and the scalar condensate are computed for the above
mentioned actions at a number of coupling values and fermion masses.
We also discuss possibilities to simulate overlap fermions dynamically
avoiding problems with low-lying eigenvalues of the overlap kernel.
\end{abstract}

\end{titlepage}

\section{Introduction}

Besides the interest in the 2-dimensional Schwinger 
model \cite{Schwinger:1962tp} in its own right 
as a quantum field theory, it
can be considered as a test laboratory 
for new theoretical concepts and ideas that aim at 
eventual applications in more demanding situations such as lattice QCD. 
In particular, 
for numerical simulations   
the lattice Schwinger model is most suitable to perform 
test studies
since the computations are much cheaper than in four dimensions 
and precise results at many     
parameter values can be obtained, see e.g. 
refs.~\cite{Farchioni:1998xg,Farchioni:1998xi,Gutsfeld:1999pu,Durr:2004ta}
for a selection of recent work in the Schwinger model. 
In this paper, we want to address the scaling properties of a number of 
fermion actions for $N_f=2$ flavours of dynamical fermions. 
To this end, we will compare
standard Wilson \cite{Wilson:1974sk}, 
Wilson twisted mass \cite{Frezzotti:2000nk,Frezzotti:2003ni}, 
hypercube \cite{Hasenfratz:1998bb} 
and overlap fermions \cite{Neuberger:1997fp} in 
their approach to the continuum limit. 
We will 
use throughout the paper the Wilson plaquette gauge action
with a coupling $\beta=1/a^2e^2$, with $a$ the lattice spacing and $e$ the 
dimensionful coupling.

Each of the above mentioned fermion actions has certain advantages and are 
used in 
present simulations of lattice QCD. Understanding  
the scaling properties when using different lattice actions and to
check the expected scaling behaviour as a function of the 
lattice spacing $a$ is certainly one of the most important 
questions in lattice calculations. However, if we think of 
dynamical fermion simulations in lattice QCD a scaling analysis is, at least
nowadays, far too computer time consuming to be addressed, 
see e.g. ref.~\cite{Jansen:2003nt} and ref.~\cite{Urbach:2005ji} for 
recent estimates of the simulation costs in lattice QCD. 
On the other hand, for the 2-dimensional Schwinger model, such simulations
are perfectly possible and give important insight into the properties of the
above mentioned actions. 

In order to study the scaling behaviour, 
we will fix the scaling variable $z\equiv (m_f\sqrt{\beta})^{2/3}$,
where $m_f$ is the fermion mass in lattice units.                
We have chosen $z=0.2,0.4,0.8$.
The fermion mass is determined from the PCAC relation where we employ 
local as well as conserved currents.  
At each of these fixed values of $z$ we compute the pseudo 
scalar mass and the scalar condensate
and follow its behaviour with decreasing value of the lattice spacing.
Performing finally a continuum limit of our results allows us to compare
to analytical predictions that are available from approximations of the 
massive Schwinger model which cannot be solved exactly. 

The paper is organized as follows. In section~2 we give the definition 
of the Wilson, hypercube, maximally twisted mass and overlap fermion operators
which 
we have employed in this work. Section~3 is devoted to the observables 
we have used and provides a description of  
the numerical simulations, in particular our attempts to simulate overlap 
fermions dynamically.   
Section~4 contains our results and section~5 the conclusions. 

\section{Lattice fermions}

In this section we give the definitions of the different kind of fermions
we have used, i.e. Wilson, hypercube, Wilson twisted mass and overlap 
fermions. Throughout the work we have employed only the Wilson plaquette
gauge action with gauge fields $U_{n,\mu}\in U(1)$, $n$ denoting a 
lattice point and $\mu=1,2$ a direction of the 2-dimensional lattice
\be
S_G[U] = \beta \sum_P\frac{1}{2}\big(U_P+U_P^\dagger\big)
\ee
with $U_P$ denoting the plaquette
\be
U_{P} = U_{n,1} U_{n+\hat1,2} U^\dagger_{n+\hat2,1} U^\dagger_{n,2} \;,
\ee
where $\hat1$ and $\hat2$ denote shifts in direction $1$ or $2$ respectively.
The coupling multiplying the plaquette gauge action is $\beta=1/g^2$. 
Denoting by $e$ the physical coupling, 
dimensions are introduced by $\beta=1/a^2e^2$.
We restrict ourselves to $N_f=2$ flavours of dynamical fermions. 

\subsection{Wilson fermions}
As a first action that can serve as a kind of benchmark action, we have 
chosen standard Wilson fermions \cite{Wilson:1974sk} with  
the Wilson operator 
$D_\mathrm{W}$
given by 
\be
\label{wilson}
D_\mathrm{W}^{n,m}= (m_0 + 2r) \delta_{n,m} -
      \frac{1}{2}\sum_\mu[(r-\sigma_\mu) U_{n,\mu} 
      \delta_{n,m-\hat\mu} +
      (r+\sigma_\mu)U^\dag_{m,\mu} \delta_{n,m+\hat\mu} ]\; . 
\ee
Here the sum goes over the two directions $\mu=1$ and $\mu=2$ and
we use the standard Pauli-matrices $\sigma_\mu, \mu=1,2,3$ with 
$\sigma_3=\mathrm{diag}(1,-1)$, $n,m$ denote lattice points with 
space and time coordinates $(x,t)$ and we suppress the Dirac indices. 
The Wilson parameter $r$ is chosen to be $r=1$ throughout the paper.
The Wilson action is expected to lead to large discretization errors in 
physical observables linear in the lattice spacing $a$ and 
hence approaches the continuum limit rather slowly. 
We will compare results obtained with the Wilson action to 
corresponding results from actions where these $O(a)$ lattice artefacts
are expected to vanish and changed to an $O(a^2)$ behaviour. 

\subsection{Hypercube fermions}

Perfect actions
\cite{Hasenfratz:1998bb} are completely free of lattice artefacts.
However, since such actions cannot be realized in practice, truncated
versions are used which are expected to inherit many of the properties
of the perfect actions. 
For this work, we have chosen a particular ansatz, the hypercube 
action   
\cite{Bietenholz:1995cy}.  
In this approach the interaction among fields on the lattice is extended
from a purely nearest neighbour interaction to a hypercube of the 
lattice. 

A general ansatz for the hypercube operator is 
\bearr
D^{n,m}_\mathrm{hyp}=
\sum_\mu \rho^\mu_{m-n} \sigma_\mu + \lambda_{m-n} \mathbbm{1} ,
\eearr
where $\rho^\mu_{m-n}$ and $\lambda_{m-n}$ are real parameters, the values of 
which have to be optimized according to some criterion, e.g. 
by demanding an improved continuum limit behaviour of physical 
observables. 

Furthermore, symmetry requirements of the action leads, in the case of
the Schwinger model, to a reduction 
of these couplings to only 
5 free parameters that will depend on the 
gauge coupling constant $g$ and the bare fermion mass $m_\mathrm{hyp}$. 
The values of $\rho$ and $\lambda$ are determined for free fermions, setting
$g=0$ and then they are taken over to the interacting case. 
As an alternative, the values of $\rho$ and $\lambda$ may be determined
by requiring that the violations of the Ginsparg-Wilson relation
are minimized. 
In this work we will, however, only work with the ``scaling optimized'' 
set of
parameters as given in ref.~\cite{Bietenholz:1999km} since we are 
interested mainly in the scaling behaviour. 
We give in the following table the values of the parameters that we have used
in this work. 

\begin{center}
\begin{tabular}{|l|l|}
\hline
$\lambda_0$ & $+1.5 + m_\mathrm{hyp}$ \\
\hline
$\lambda_1$ & $-0.25$ \\
\hline
$\lambda_2$ & $-0.125$ \\
\hline
$\rho^{(1)}$ & $+0.334$\\
\hline
$\rho^{(1)}$ & $+0.083$\\
\hline
\end{tabular}
\end{center}

The full hypercube Dirac operator is given by 
\bearr
\label{hypercube}
D_{\mathrm{hyp}}&=& \lambda_0 + H^{(1)}+H^{(2)} \qquad \nonumber\\ 
(H^{(1)})^{nm}&=& \sum_{\mu=1}^2 \Big(U_{n,\mu} \delta_{n,m-\hat \mu} (\lambda_1 + \sigma_\mu \rho^{(1)})
         + U^\dag_{n-\hat\mu,\mu} \delta_{n,m+\hat\mu} (\lambda_1 - \sigma_\mu \rho^{(1)}) \Big) \nonumber\\
(H^{(2)})^{nm}&=&  
       \frac{1}{2} \Big( (U_{n, 1} U_{n+\hat 1,2} + U_{n, 2}U_{n+\hat 2, 1} ) \,
                    \delta_{n,m-\hat 1-\hat 2} (\lambda_2 + (\sigma_1+\sigma_2) \rho^{(2)}) \nonumber \\
  && \ \ + (U_{n, 1} U^\dag_{n+\hat 1-\hat2,2}
                                      +U^\dag_{n-\hat2,2} U_{n-\hat2,1} ) \,
                    \delta_{n,m-\hat1+\hat2} (\lambda_2 + (\sigma_1-\sigma_2) \rho^{(2)}) \\
  && \ \ + (U^\dag_{n-\hat1,1}U_{n-\hat1,2}
                                      + U_{n,2}U^\dag_{n+\hat2-\hat1,1}) \,
                    \delta_{n,m+\hat1-\hat2} (\lambda_2 + (-\sigma_1+\sigma_2) \rho^{(2)}) \nonumber \\
  && \ \ + (U^\dag_{n-\hat1,1}U^\dag_{n-\hat1-\hat2,2}
                                      +U^\dag_{n-\hat2,2}U^\dag_{n-\hat1-\hat2,1}) \,
                    \delta_{n,m+\hat1+\hat2} (\lambda_2 + (-\sigma_1-\sigma_2) \rho^{(2)})\Big)\; . \nonumber
\eearr

\subsection{Twisted mass fermions}

The twisted mass formulation of lattice fermions has been introduced
originally to regulate small, unphysical eigenvalues of the 
Wilson lattice Dirac operator \cite{Frezzotti:2000nk}. 
In order to keep an $O(a)$ improvement, the twisted mass setup has been
first developed in the $O(a)$ Symanzik improved theory. 
However, it then was realized that by a 
careful tuning of the parameters of the Wilson twisted mass action, 
an automatic, full $O(a)$-improvement can be reached, 
leading to lattice discretization
errors that appear only in $O(a^2)$ and hence allow for a much 
accelerated continuum limit \cite{Frezzotti:2003ni}.
The Wilson twisted mass formulation has received a lot of attention
recently and a number of tests in the quenched 
\cite{Jansen:2003ir,Abdel-Rehim:2005gz,Jansen:2005gf,Jansen:2005kk,Jansen:2005cg}
and partly also 
for full dynamical fermions 
\cite{Farchioni:2004fs,Farchioni:2004ma,Farchioni:2004us,Jansen:2005tu} 
has been performed.  

In order to introduce twisted mass fermions, let us start with 
the continuum, Euclidean action,
\bearr
S[\bar\psi,\psi]=\int d^2x \bar\psi (\sigma_\mu D_\mu + m + i \mu_f \sigma_5 \tau_3) \psi ,
\eearr
where $\tau_3$ is the third Pauli matrix, acting in 
flavour space and 
$\mu_f$ represents the twisted mass parameter.
The transformation  
\bearr
\label{tmtransformation}
\psi' = e^{i \omega \sigma_5 \frac{\tau_3}{2}} \psi \nonumber \\
\bar\psi'=\bar\psi e^{i \omega \sigma_5 \frac{\tau_3}{2}}
\eearr
leaves this form of the action invariant with rotated mass parameters, 
\bearr
m'&=&m \cos\omega + \mu_f \sin\omega \nonumber \\
\mu_f'&= -& m \sin\omega + \mu_f \cos\omega\; ,
\eearr
with  the rotation (``twist'') angle $\omega$,  
\bearr
\label{transformationangle}
\tan\omega = \frac{\mu_f}{m} \; .
\eearr

On the lattice, the Wilson term is not invariant under the rotation
in eq.~(\ref{tmtransformation}) and the twisted mass operator 
takes the form
\bearr
\label{tmaction}
D_\mathrm{tm}=D_\mathrm{W}^0 +  \big[ m_0 + i \mu_f \sigma_5 \tau_3 \big] 
\eearr
where $D^0_\mathrm{W}$ denotes the 
Wilson operator without the mass term, i.e. 
\be
D_\mathrm{W}^0=D_{\mathrm{W}}[m_0=0]\; . 
\label{d0}
\ee

It can be shown that the twisted mass action leads to an O(a)-improvement
when the angle $\omega=\pi/2$. It goes beyond the scope of this paper
to provide the arguments for this remarkable result the derivation of 
which can be found in ref.~\cite{Frezzotti:2003ni}. 
We only would like to remark that 
in order 
to obtain this value of the twist angle is equivalent to tune the 
bare fermion mass parameter $m_0$ to a critical value $m_0^\mathrm{crit}$. 

One very important aspect of twisted mass fermions is that 
there is a particular definition of $m_0^\mathrm{crit}$ from 
the vanishing of the PCAC fermion mass that does not only lead to an 
$O(a)$-improvement 
\cite{Aoki:2004ta,Sharpe:2004ny,Abdel-Rehim:2005gz,Frezzotti:2003ni,Jansen:2005gf}
but also substantially reduces cut-off effects that 
appear in $O(a^2)$ as has been demonstrated in 
\cite{Frezzotti:2005gi,Jansen:2005kk}.
In this paper, we will use a value of $m_0^\mathrm{crit}$ that was 
obtained in the pure Wilson fermion theory without twisted mass term from
the PCAC relation. 
The value of $m_0^\mathrm{crit}$ was then tuned in such a way that 
the corresponding PCAC fermion mass vanishes.
After having determined 
$m_0^\mathrm{crit}$ in this way, we 
varied the twisted mass parameter $\mu_f$ to realize the fermion and 
pion masses we are interested in. 


\subsection{Overlap fermions}

A lattice Dirac operator $D_{\rm GW}$ that satisfies the
Ginsparg-Wilson relation \cite{Ginsparg:1981bj}
\be  \label{GWR}
D_{\rm GW} \gamma_{5} + \gamma_{5} D_{\rm GW} = 2 a D_{\rm GW} \gamma_{5} R
D_{\rm GW} \ ,
\ee
where $R$ is a local term, leads to an action that has an 
exact (lattice) chiral symmetry eliminating thus automatically
$O(a)$ cut-off effects.
The realization of an operator $D_{\rm GW}$ that we use here is the overlap
fermion, which is characterized by the overlap Dirac operator 
\cite{Neuberger:1997fp}.
For $R_{n,m}= \frac{\delta_{n,m}}{2\rho}$ it takes the form
\bea
D_{\rm ov} &=& \Big( 1 - \frac{m_{\rm ov}\bar a}{2} \Big) D_{\rm ov}^{(0)}
+ m_{\rm ov} \ ,
\label{overlapWithMass}\\
D_{\rm ov}^{(0)} &=& \frac{1}{\bar a}
\left\{ 1 + D_0 / \sqrt{ D_0^{\dagger} D_0} \right\}
\label{overlap}
\eea
where $D_0$ is the so-called overlap kernel operator and 
$\bar a\equiv a/\rho$. For  
the kernel 
$D_0$, there is a large choice. In the following, however, we will
only use the hypercube operator, i.e. $D_0=D_\mathrm{hyp}$
with fixed $\lambda_0= 0.5$, i.e. setting
$m_\mathrm{hyp}=-1$.
This corresponds to
setting the parameter $\rho=1$ in eq.~(\ref{overlap}) and guarantees
the locality of the overlap operator 
\cite{Hernandez:1998et}.

We remark that we have realized the square root operator 
$1/\sqrt{ D_0^{\dagger} D_0}$ in eq.~(\ref{overlap}) by Chebyshev 
polynomials $P_{n,\epsilon}$ with degree $n$ and a lower bound 
$\epsilon$. This Chebyshev polynomial shows an exponential convergence 
rate in the interval $[\epsilon,1]$. Setting $\epsilon$ to the lowest 
eigenvalue of the overlap kernel operator 
$D_0^{\dagger} D_0$ which is normalized to one,  
we always have chosen the degree $n$ such that we reach 
machine precision for the evaluation of $1/\sqrt{ D_0^{\dagger} D_0}$.
We used eigenvalue deflation and projected out a number of 
low-lying eigenvalues. 
In this case, $\epsilon$ was set to the lowest non-projected 
eigenvalue of $D_0^{\dagger} D_0$.


\section{Observables and simulations}

In this section we give the operators that we have 
used to determine the physical observables we have computed and describe 
the numerical simulations. In particular, we discuss some of our attempts
to perform dynamical overlap simulations avoiding problems with 
very low-lying eigenvalues of the overlap kernel operator.

\subsection{Observables}
\vspace*{0.3cm}
\noindent {\bf 2-point functions}
\vspace*{0.3cm}

\noindent Generally, the bi-linear operators are given by
\bearr
\label{biloperator}
\mathcal{\tilde O}_n^{\Gamma \tau}=\bar\psi_{n} \Gamma \tau \psi_{n} ,
\eearr
where $\Gamma$ stands for certain combinations of the Pauli matrices
$\sigma$ acting in Dirac-space, while the Pauli matrices  
$\tau$ act in flavour space.
We will consider the following operators, 
\bearr
\label{pseudoscalar}
\mathcal{\tilde P}^a_n &=& \bar\psi_n \sigma^{(Dirac)}_3 
  \tau_a^{(flavour)} \psi_n \\
\label{scalar}
\mathcal{\tilde S}_n &=& \bar\psi_n  
  \psi_n \\
\label{pseudovector}
\mathcal{\tilde A}^{\mu a}_n &=& \bar\psi_n \sigma^{(Dirac)}_3 
\sigma^{(Dirac)}_\mu \tau_a^{(flavour)} \psi_n \\
\label{vectorcurrent}
\mathcal{\tilde V}^{\mu a}_n &=& \bar\psi_n \sigma^{(Dirac)}_\mu 
\tau_a^{(flavour)} \psi_n ,
\eearr
with $\mathcal{\tilde P}$ the pseudo scalar, $\mathcal{\tilde S}$ the scalar,
$\mathcal{\tilde A}$ the axial-vector and $\mathcal{\tilde V}$ the vector
operator.

For the twisted mass fermions, the operators take a modified form
as can be obtained from the field transformation according to 
eq.~(\ref{tmtransformation}), leading to 
\bearr
\mathcal{\tilde P}^{a (ph)}_n &=& \left\{
\begin{array}{ll}
\mathcal{\tilde P}^{a (tb)}_n & \textrm{if} \ a=1,2\\
\cos\omega \, \mathcal{\tilde P}^{3 (tb)}_n + \sin\omega \, 
\mathcal{\tilde S}_n^{(tb)} \hspace{0.97cm}
  & \textrm{if} \  a=3
\end{array} \right.  \\
\mathcal{\tilde A}^{\mu a (ph)}_n &=& \left\{
\begin{array}{ll}
\cos\omega \, \mathcal{\tilde A}^{\mu a (tb)}_n
                       + \sin\omega \, \epsilon_{ab} \mathcal{\tilde V}^{\mu b (tb)}_n
  & \textrm{if} \  a=1,2\\
\mathcal{\tilde A}^{\mu 3 (tb)}_n
  & \textrm{if} \  a=3
\end{array} \right.  \\
\mathcal{\tilde V}^{\mu a (ph)}_n &=& \left\{
\begin{array}{ll}
\cos\omega \, \mathcal{\tilde V}^{\mu a (tb)}_n
                       + \sin\omega \, \epsilon_{ab} \mathcal{\tilde A}^{\mu b (tb)}_n
  & \textrm{if} \  a=1,2\\
\mathcal{\tilde V}^{\mu 3 (ph)}_n .
  & \textrm{if} \  a=3
\end{array} \right.
\eearr
Here ``ph'' denotes the physical and ``tb'' the twisted basis. 
Note that 
in the special case $\omega=\pi/2$ for $a=1,2$,  
$\mathcal{\tilde V}$ and $\mathcal{\tilde A}$ just interchange their role 
while $\mathcal{\tilde P}$ remains invariant. 
In particular, the scalar operator is given by the 3rd component of the 
pseudo scalar operator.

For the Wilson, hypercube and Wilson twisted mass  
fermions we computed the 2-point correlation functions by standard 
techniques using a conjugate gradient solver. For overlap fermions we 
calculated the correlators both from using a conjugate gradient solver
and from 
eigenvectors $\phi_i$ and eigenvalues $\lambda_i$ of the
overlap operator in eq.~(\ref{overlapWithMass}). 
A generic non-singlet 
correlator $C(n,m)$ is then expressed in terms of the fermion 
propagator $S(n,m)$ and a suitable Dirac structure $\Gamma$, 
corresponding to the operators listed above, as
\bearr
C(n,m)&=&tr[\Gamma S(n,m)\Gamma S(m,n)] \nonumber \\
&=&\sum_{\lambda_{i},
\lambda_{j}}\frac{1}{\lambda_{i}\lambda_{j}}\sum_{\alpha\beta\gamma\delta}
    \bigg[(\phi^{\dag\alpha}_{j}(n)\Gamma_{\alpha\beta}\phi^{\beta}_{i}(n))
(\phi^{\dag\gamma}_{i}(m)\Gamma_{\gamma\delta}
      \phi^{\delta}_{j}(m))\bigg] \; .
\label{correv}
\eearr

\vspace*{0.3cm}
\noindent {\bf PCAC fermion mass}
\vspace*{0.3cm}

The bare mass parameter for Wilson ($m_0$) and hypercube 
($m_\mathrm{hyp}$) fermions receives 
an additive mass renormalization, necessitating the determination 
of a critical fermion mass $m_\mathrm{crit}$ which we computed 
by the vanishing of the PCAC fermion mass, extracted from 
the 
PCAC relation 
\bearr
\langle \partial_\mu \mathcal{\tilde A}^{\mu} (n) \mathcal{\tilde O} \rangle
         = 2 m \langle \mathcal{\tilde P}(n) \mathcal{\tilde O} \rangle .
\eearr
Choosing  
$\mathcal{\tilde O}=\mathcal{\tilde P}$ and projecting to zero momentum
(denoting by $\mathcal{A}$ and $\mathcal{P}$ the currents summed over space), 
we arrive at 
\begin{equation}
\label{pcacmass}
m_f(t) = \frac{\langle \partial_2 \mathcal{A}^{2}_t \mathcal{P}_0 \rangle}{2 \langle \mathcal{P}_t \mathcal{P}_0 \rangle} .
\end{equation}

For the currents that appear in eq.~(\ref{pcacmass}) one may use the 
local currents of eqs.~(\ref{pseudoscalar}-\ref{vectorcurrent}). 
An alternative is to employ also conserved currents. 
A very general way to derive the conserved currents 
is given in \cite{Hasenfratz:2002rp}.
The vector and axial currents are then given by 
\bearr
\label{conservedVectorCurrent}
\mathcal{\tilde V}^{\mu a}_n&=&\tilde J^{\mu R}_n+\tilde J^{\mu L}_n \\
\label{conservedAxialCurrent}
\mathcal{\tilde A}^{\mu a}_n&=&\tilde J^{\mu R}_n-\tilde J^{\mu L}_n
\eearr
where 
\bearr
\tilde J^{\mu R}_n=\frac{1}{2} \bar\psi (1-\sigma_3) \mathcal{K}^{\mu, n}(1+\sigma_3) \psi \\
\tilde J^{\mu L}_n=\frac{1}{2} \bar\psi (1+\sigma_3) \mathcal{K}^{\mu, n}(1-\sigma_3) \psi
\eearr
and 
\bearr
\mathcal{K}^{\mu, n} = - i \frac{\delta D(U^{(\alpha)})}{\delta \alpha_{n,\mu}}\Big|_{\alpha=0}\;,\;\;
\qquad U^{(\alpha)}_{n,\mu}=e^{i\alpha_{n,\mu}} U_{n,\mu}
\eearr
where $D(U^{(\alpha)}_\mu)$ denotes the lattice Dirac operator 
used. 
This method of constructing the conserved currents
is very useful when complicated lattice Dirac operators are
considered since their construction from the current conservation
condition can be very cumbersome.
We followed this prescription to compute the conserved currents,
only for the overlap fermions we used the local point currents. 
We remark that for the twisted mass and the overlap case in principle 
also the 
bare fermion masses, $\mu_f$ and $m_\mathrm{ov}$ can be used. However, 
since this could lead to very different lattice artefacts, 
we employed also in these cases the 
fermion mass derived from the PCAC relation for all physical results 
presented in the following in order to be able to directly compare 
the different fermion actions.

\vspace*{0.3cm}
\noindent {\bf Scalar condensate} 
\vspace*{0.3cm}

Besides the 2-point functions which will provide the pseudo scalar mass, 
we considered also the scalar condensate, 
$\Sigma\equiv \langle \bar\psi \psi\rangle$. A first method
to compute $\Sigma$ is by calculating $\mathrm{Tr}D^{-1}$  
using Gaussian noise sources. We will denote the so computed values of $\Sigma$
as $\Sigma_\mathrm{direct}$. 
This quantity develops 
in the 
case of Wilson and hypercube fermions 
a divergent piece $\propto 1/a$.  
%
A second way, which avoids the appearance of the divergent piece from the
beginning,
 is to use the integrated axial
Ward identity leading to a ``subtracted'' scalar condensate, 
$\Sigma_\mathrm{sub}$ 
\cite{Bochicchio:1985xa}. 
We remark that in the case of overlap fermions
we computed $\Sigma$ from the improved scalar operator 
$\langle \bar\psi (1-\frac{a}{2}D)\psi\rangle$ which we evaluated from 
the
the eigenvalues of the overlap operator. 
More precise definitions and a further discussion will be provided 
in section~4.

\subsection{Simulations}

In our work, we have used for the Wilson, the hypercube and the twisted mass
fermion action a standard Hybrid Monte Carlo algorithm (HMC) 
\cite{Duane:1987de}.
In the case of overlap fermions a straightforward implementation of the 
HMC algorithm is not suitable since  
the variation of the Hamiltonian with 
respect to the gauge field can lead to very large forces proportional 
to $(D_0^\dagger D_0)^{-3/2}$ when the overlap kernel operator $D_0$  
develops exceptionally small eigenvalues. 
See refs.~\cite{Fodor:2003bh,Cundy:2005pi,DeGrand:2005vb} for variants 
of the HMC algorithm 
that circumvent this difficulty. 

For the overlap fermion results we used 
configurations 
that were generated in the 
pure gauge theory only and performed a reweighting with the 
overlap fermion determinant to obtain physical observables for
dynamical $N_f=2$ flavours of fermions.
However, 
we also tried several possibilities 
to avoid the problem with low-lying eigenvalues 
of the overlap kernel operator $D_0$ by replacing $D_\mathrm{ov}$ 
by some operator $D_\mathrm{ov}^\mathrm{approx}$ that is a good 
approximation to $D_\mathrm{ov}$ but which  
is safe against these low-lying 
eigenvalues. The simulation can then be made exact 
again by adding a correction step employing the ratio 
$D_\mathrm{ov}/D_\mathrm{ov}^\mathrm{approx}$. 

The general idea is to write (we use for simplicity only a single 
operator here, in practice one would have to use the operator 
$D^\dagger D$, of course)
\be
\mathrm{det}D_\mathrm{ov}=\mathrm{det}D_\mathrm{ov}^\mathrm{approx}\cdot
\mathrm{det}\left[\frac{D_\mathrm{ov}}{D_\mathrm{ov}^\mathrm{approx}}
\right]
\equiv \mathrm{det}D_\mathrm{ov}^\mathrm{approx}\cdot \R\; .
\ee
While $\mathrm{det}D_\mathrm{ov}^\mathrm{approx}$ would be used
in the HMC algorithm, 
the remaining determinant ratio $\R$ could be implemented either in an
additional 
accept/reject step or it could be included as a reweighting factor in the 
computation of a given observable.

A crucial question in such an approach is, whether an operator 
$D_\mathrm{ov}^\mathrm{approx}$ can  be found such that the 
fluctuations in $\R$ are small enough to obtain statistically 
significant results. 
We decided therefore to test 
this idea by computing $\R$ stochastically using 
$n=10$ Gaussian noise vectors  
on a number of gauge field configurations 
generated in the pure gauge theory at $\beta=3$ on a $16^2$ lattice. 
In the following we will describe the results of these tests for three
choices of $D_\mathrm{ov}^\mathrm{approx}$.

\vspace*{0.3cm}
\noindent {\bf Case of $D_\mathrm{ov}^\mathrm{approx}=D_\mathrm{hyp}$}
\vspace*{0.3cm}

As a first trial, we used the hypercube operator $D_\mathrm{hyp}$ as an
approximation to $D_\mathrm{ov}$. This choice has been motivated by the
fact that the operator $D_\mathrm{hyp}$ is constructed to be approximating
the Ginsparg-Wilson relation, resulting in a very similiar eigenvalue
spectrum for both operators \cite{Bietenholz:1999km}.
Note also that we have used the hypercube
operator itself as an overlap kernel operator $D_0$. 
Let us remark that we re-write the massive overlap operator 
as $D_\mathrm{ov}^{(0)} + m_\mathrm{ov}/(1- m_\mathrm{ov}/2)$  
in order to better match the eigenvalue spectra of both operators. 

We computed $\R$ as a function of the hypercube bare fermion mass
$m_\mathrm{hyp}$ in a range $-0.25<m_\mathrm{hyp}<0.25$. 
However, we found that the fluctuations of $\R$ were extremely large for
all values of $m_\mathrm{hyp}$ we have tested.
Thus we had to conclude that $D_\mathrm{hyp}$ cannot serve as an 
infrared safe, approximate operator for the overlap simulations when 
$R_\mathrm{det}$ is used in a stochastic correction step.
We cannot exclude, of course, that by using an improved hypercube operator,
e.g. by adding the clover term or performing smearing of the link variables,
the situation could be improved. 
However, given the negative findings of our investigation, 
we did not pursue this direction. 
What we tried instead, is to use $D_\mathrm{hyp}$ as the guidance 
Hamiltonian in the molecular dynamics part of the 
Hybrid Monte Carlo algorithm while keeping 
$D_\mathrm{ov}$ as the exact overlap operator for the accept/reject 
Hamiltonian. Amazingly, despite the negative results described here
for $R_\mathrm{det}$, we found 
that this led to reasonable acceptance rates, at least for not too large
systems, see ref.~\cite{Christian:2005gd} for a further discussion of this point.

\vspace*{0.3cm}
\noindent {\bf Modified Chebyshev polynomial}
\vspace*{0.3cm}

As a second attempt, we tried to modify the range of the Chebyshev 
polynomial employed in the construction of the overlap operator. 
There, the square root of the kernel is computed by a Chebyshev 
polynomial $P_{n,\epsilon}$ of degree $n$ in the interval 
$\left[\epsilon,1\right]$,          
\be
P_{n,\epsilon}(D_0^\dagger D_0) \approx 1/\sqrt{D_0^\dagger D_0}\; .
\label{pn}
\ee
We have chosen the approximation accuracy in eq.~(\ref{pn}) to be 
very high, compatible with machine 
(64-bit) precision. 
The value of $\epsilon$ is chosen to correspond to the lowest 
eigenvalue of $D_0^\dagger D_0$. 
In principle, it is possible to use instead of the ``exact'' polynomial 
with parameters $n$ and $\epsilon$ a modified polynomial with 
parameters ${\tilde n}<n$ and $\tilde{\epsilon} > \epsilon$. 
Clearly, if $\tilde{\epsilon}$ could be chosen to be well above 
the smallest eigenvalues of $D_0^\dagger D_0$ then using an overlap 
operator with a polynomial $P_{{\tilde n},\tilde{\epsilon}}$ in the molecular 
dynamics part would lead to an infrared safe simulation. 

As a start situation, we had chosen 
$\tilde{\epsilon}=\epsilon$ and reduced only the degree of the polynomial 
successively. We found that it is indeed possible to reduce $n$ 
substantially to
${\tilde n}\approx 1/3\cdot n$ without having large fluctuations in the 
determinant ratio 
$\R$. This is a very positive outcome since it gives rise to a
substantial acceleration of
simulations with the overlap operator. 
However, when we tried to change the value of $\tilde\epsilon$ only 
slightly, we observed
immediately large fluctuations in $\R$. 
For $\tilde{\epsilon}=1.1\cdot\epsilon$
the fluctuations in $\R$ became already so large that there is no chance 
to obtain a statistically significant result in such a setup.

\vspace*{0.3cm}
\noindent {\bf Explicit infrared regularization}
\vspace*{0.3cm}

As a third (and last) attempt, we tried to use an approximate overlap 
operator that has an infrared regulator built in. In particular, we 
studied the situation where we use
a modified sign function
\be
D_0/\sqrt{D_0^\dagger D_0} \rightarrow D_0/(\sqrt{D_0^\dagger D_0+\delta)}\; .
\label{signmod}
\ee
The (optimistic) expectation here is that by choosing $\delta$ large enough, 
without having too large fluctuations in the determinant ratio $\R$, 
the overlap operator with such a modified sign function would be infrared safe
for the HMC simulation.

\begin{figure}[tbp]
  \centering
  \includegraphics[height=12.0cm]{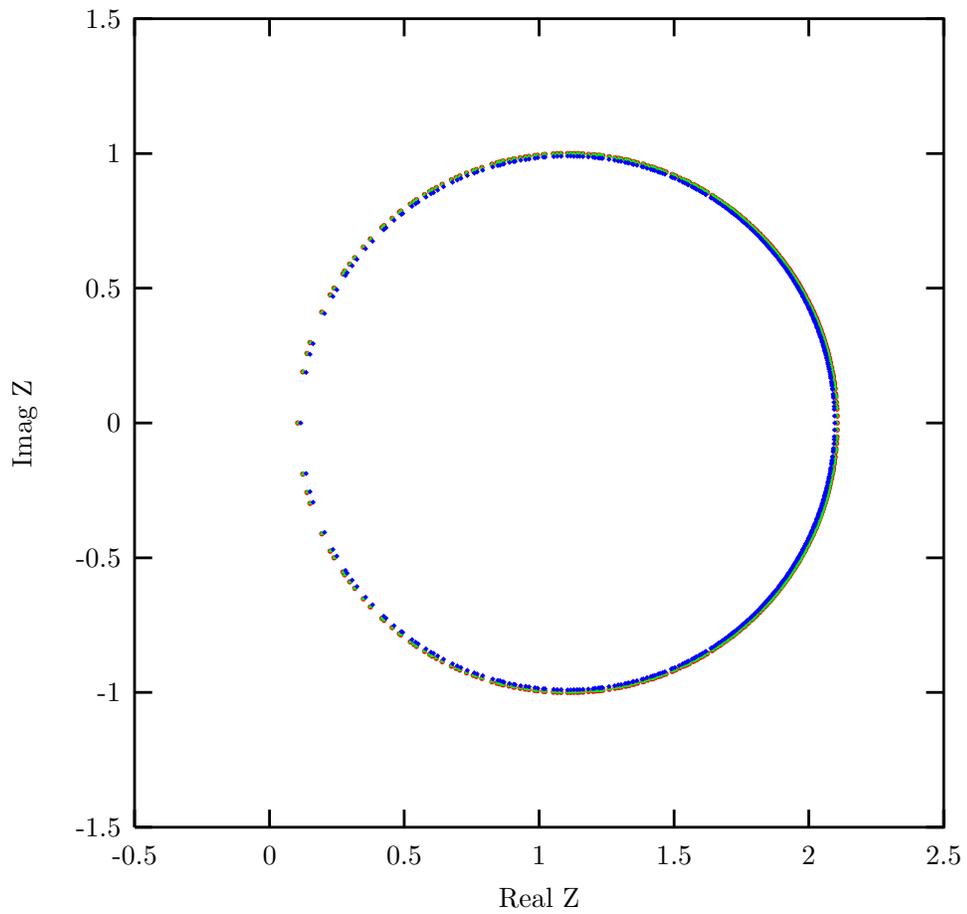}
  \caption{The eigenvalues of the exact overlap operator and a modified
overlap operator when the sign function is changed according 
to eq.~(\ref{signmod}). The plot contains the eigenvalues for 
$\delta=0$ (the original
overlap operator) and for $\delta=0.0001,0.001,0.01$. For all these cases, 
the eigenvalues are so close to each other that they cannot be 
distinguished in the graph.}
  \label{fig:circle}
\end{figure}

Such an optimistic expectation is not completely unfounded 
when one inspects the eigenvalues of 
the exact overlap operator $D_\mathrm{ov}$ and the sign function modified
one, $D_\mathrm{ov}^\mathrm{approx}$. Indeed, we show in fig.~\ref{fig:circle}
the eigenvalues when choosing $\delta=0,0.0001,0.001,0.01$ for a typical configuration.

\begin{figure}[tbp]
  \centering
  \includegraphics[height=11.0cm,width=14.0cm]{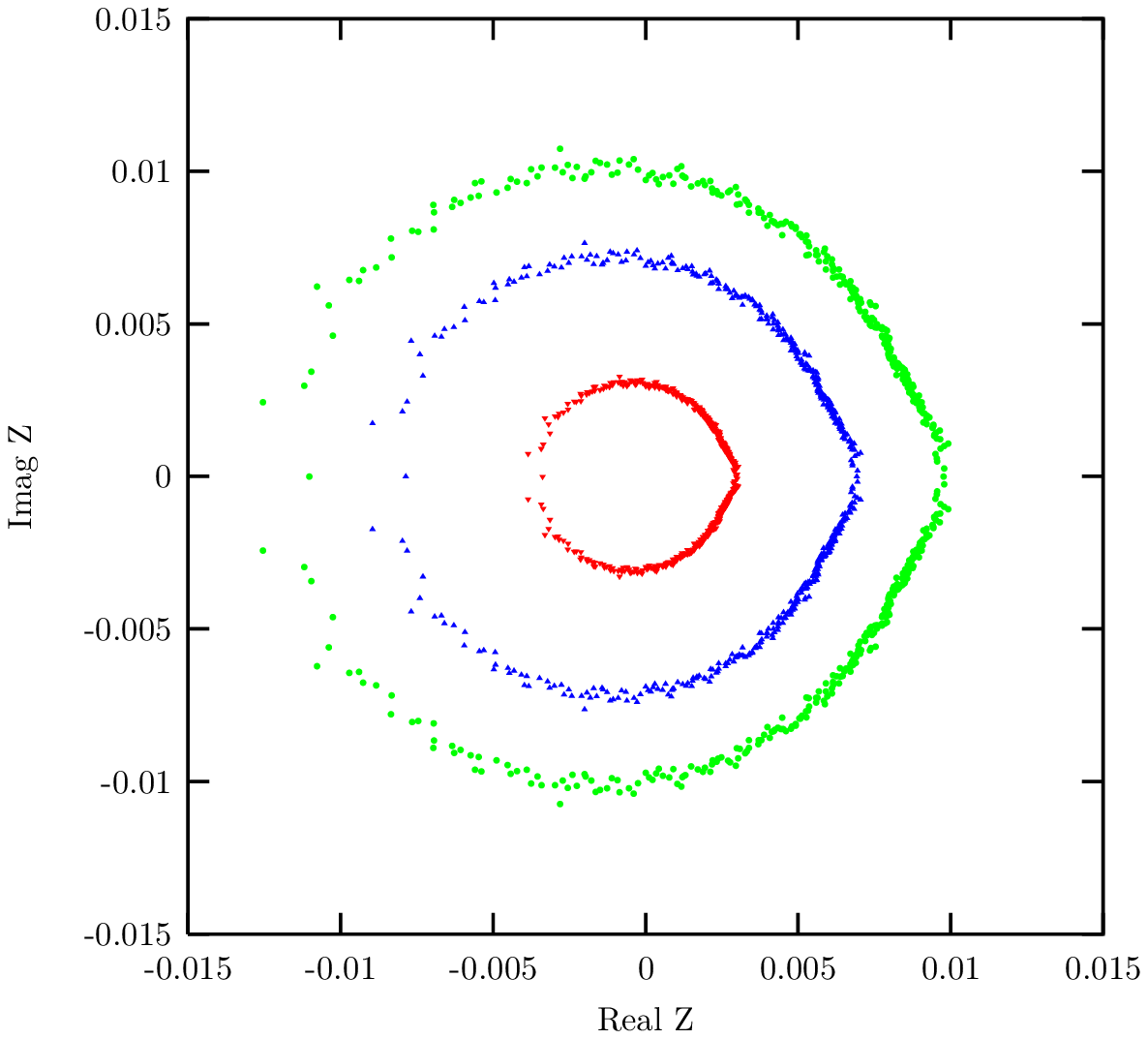}
  \caption{The difference spectra of the overlap operator with 
$\delta=0$ and the operators with $\delta=0.01$ (most outer spectrum),
$\delta=0.001$ (middle spectrum) and $\delta=0.0001$ (most inner spectrum). 
The difference spectra for $\delta=0.001$ and $\delta=0.0001$ have been 
scaled with a factor $7$ and $30$, respectively, in order to be able 
to include them in
the plot.}
  \label{fig:lemons}
\end{figure}

As can be seen,  
all the eigenvalues lie on the expected circle
and a difference for various choices of $\delta$ is not visible. 
In fig.~\ref{fig:lemons} we show the difference of the eigenvalues between 
the spectra using $\delta=0$ and $\delta\ne 0$. In building the 
difference of the eigenvalues we used an ordering of the eigenvalues 
with respect to their real part. The scale in the plot is 
chosen for the situation of $\delta=0.01$. We have rescaled the difference 
in the eigenvalues by a factor of $30$ for $\delta=0.0001$ and $7$ 
for $\delta=0.001$ in order to plot all difference spectra in one common 
plot. It appears that the difference spectra do not build a perfect circle
shape but are rather lemon shaped. Nevertheless the lemon distortion of 
the circle happens at a rather small scale of $O(10^{-2})$ for the case of 
$\delta=0.01$ while for smaller values of $\delta$ the difference becomes 
even smaller. 
Note that the smallest eigenvalue is 
$\lambda_\mathrm{min}(D_0^\dagger D_0)=0.24$
for this configuration. 

\begin{figure}[tbp]
  \centering
\includegraphics[width=12.0cm]{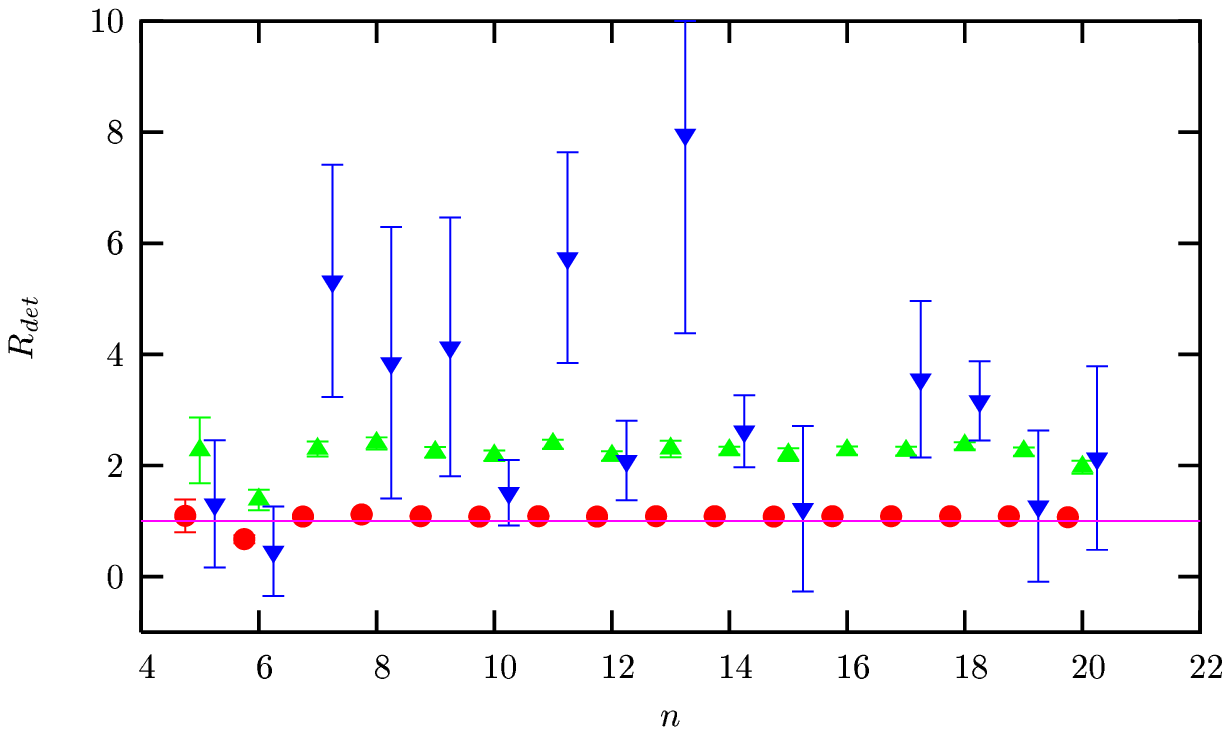}
  \caption{The determinant ratio $\R$ as a function of the degree $n$ 
of the Chebyshev polynomial. We plot the cases for $\delta=0.0001$
(circles), $\delta=0.001$ (upward triangles) and 
$\delta=0.01$ (downward triangles).}
  \label{fig:fluctuations}
\end{figure}

The smallness of the distortions in the spectrum as observed in 
fig.~\ref{fig:lemons} appears to be quite 
promising for this choice of an approximate overlap operator. 
However, the effects on the fluctuations are considerable as can be seen in
fig.~\ref{fig:fluctuations}. For a value of $\delta=0.0001$ 
(circles) 
the determinant 
ratio $\R\approx 1$ and has very small fluctuations until the degree of the 
polynomial is lowered to ${\tilde n}=8$ which is a factor of three smaller than 
the original degree of the polynomial. 
However, already for $\delta=0.001$ (upward triangles) 
$\R$ starts to develop some 
fluctuations and finally, for $\delta=0.01$ (downward triangles) 
the fluctuations are 
becoming so large that realistic simulations with such a value of $\delta$ 
cannot be performed.

As a conclusion, we found that none of the above described attempts 
led to a satisfactory 
solution for an infrared safe dynamical overlap simulation. 
As said above, 
we therefore have performed the overlap simulations finally 
by generating pure gauge configuration and using a reweighting with 
the determinant, computed exactly from the eigenvalues.

\vspace*{0.3cm}
\noindent {\bf Summary of simulation setup}
\vspace*{0.3cm}

In the following table we shortly summarize which technique we have used 
for the different lattice fermions for the simulations and for the 
observables. 

\begin{center}
\begin{tabular}{|l||l|l|}
  \hline
  & Wilson / tm / hypercube & overlap \\
  \hline
  \hline
  simulations & standard HMC &  determinant reweighting \\
  \hline
  operators & conjugate gradient solver & eigenvalues \\
            &                           & \qquad \& eigenvectors/ \\
            &                           & conjugate gradient solver \\
  \hline
  currents & local / conserved & local\\
  \hline
pseudo scalar correlator & $\langle\mathcal{P P}\rangle$ &  $\langle\mathcal{P P}-\mathcal{S S}\rangle$\\
  \hline
  scalar condensate & Ward-Takahashi-identity & $\langle \bar\psi (1-\frac{a}{2}D)\psi\rangle$\\
  \hline
\end{tabular}
\end{center}

Let us end this section by a small technical remark. For the computations 
of the eigenvalues and eigenvectors we used the LAPACK \cite{laug} routine. In the 
course of our work we found that this routine does not compute correctly
the eigenvectors in the case of degenerate zero modes. It rather gives 
linear combinations of the exact solutions which led to a problem 
in the computation of the pseudo scalar correlator using the 
operator $\mathcal{P}$ if configurations with topological
charge $|Q|>1$ are considered. The effect shows up in such a way that 
the pseudo scalar correlator                                       
eq.~(\ref{pseudoscalar})                                          
develops a 
plateau like behaviour for values of Euclidean time $t$ close to the middle
of the lattice.
This behaviour made it practically impossible to extract
the pseudo scalar mass and the PCAC fermion mass.
By re-diagonalizing the zero mode sector we could verify that this effect
goes away. 

As a simple solution for our simulations we decided to always take the 
correlator 
$\langle \mathcal{ P}(t)\mathcal{ P}(0)-
\mathcal{ S}(t)\mathcal{ S}(0)\rangle$ in which this zero mode
contribution is cancelled out \cite{Blum:2000kn}. 
The only exception is the case 
$z=0.8$, where the pseudo scalar masses are so high that it became         
very difficult to disentangle them from the scalar masses when the
$\langle \mathcal{ S}(t)\mathcal{ S}(0)\rangle$ was subtracted of. 
Therefore we used for this large value of $z$ the conjugate gradient solver
to compute the correlators.

\section{Results} 

It is the main aim of this paper to check the scaling behaviour of 
the fermion actions described in section 2. 
To this end, we have chosen the simulation parameters such that 
we fix the scaling variable 
\be
z\equiv\left(m_f/g\right)^{2/3}=\left(m_f\sqrt{\beta}\right)^{2/3}\; .
\label{z}
\ee 
We performed simulations for a wide range of $\beta$-values, 
$0.1\le \beta \le 6$. For Wilson, hypercube and twisted mass fermions 
our lattices were mainly of size $V=L^2=32^2$.
Only at $\beta=5,6$ and $z=0.2$ we went up to $V=48^2$ lattices. 
For the overlap fermion simulations we mainly used $V=20^2$ and,
for $z=0.2$, $V=24^2$ lattices. Only at $\beta=5$ and $z=0.2$ we used a
$V=28^2$ lattice. It is our experience that for larger lattices the 
determinant reweighting technique used for the overlap fermion simulations
are not practical since they lead to large fluctuations which spoil the 
signal to noise ratio. 
We finally mention that all our error analyzes are based on the method 
described in ref.~\cite{Wolff:2003sm} including thus the autocorrelation 
times in the computation of the errors. 

As a prime quantity we 
will test the scaling behaviour of 
$m_\pi\sqrt{\beta}$, performing finally a continuum limit for 
all actions
used. In the continuum, there are two approximate calculations 
for $m_\pi^\mathrm{cont}/e$ for the massive Schwinger model.

The first is performed by Smilga \cite{Smilga:1996pi} who finds
for strong coupling and small fermion mass 
\bearr
\label{smilgaExpansion}
\frac{m_\pi^\mathrm{cont}}{e} &\approx& 2^{5/6} \text{e}^{\gamma/3} \Big(\frac{\Gamma(3/4)}{\Gamma(1/4)}\Big)^{2/3}
                 \frac{\Gamma(1/6)}{\Gamma(2/3)}\Big(\frac{m}{e}\Big)^{2/3}\\
&=& 2.008 \Big(\frac{m}{e}\Big)^{2/3}\; . 
\eearr

\noindent For large masses, 
Gattringer in ref.~\cite{Gattringer:1995iq} finds from a semi-classical 
analysis
\bearr
\label{gattrexpansion1}
\frac{m_\pi^\mathrm{cont}}{e} &\approx& \text{e}^{2\gamma/3} \frac{2^{5/6}}{\pi^{1/6}} \Big(\frac{m}{e}\Big)^{2/3} \\
 &=& 2.163 \Big(\frac{m}{e}\Big)^{2/3} \; .
\eearr
Note that we give these results in terms of the dimensionful quantities, 
coupling $e$, 
fermion mass $m$ and pseudo scalar mass $m_\pi^\mathrm{cont}$. 
Both expression are approximate computations and it is interesting to 
compare these against our non-perturbative calculations.

For completeness, we also give here analytical expressions for the 
scalar condensate as available in the literature. The first is again 
from ref.~\cite{Smilga:1996pi}, 
\be
\label{anacond}
\frac{\Sigma^\mathrm{cont}}{e} \approx 0.388 \Big(\frac{m}{e}\Big)^{1/3}\; .
\ee
A second expression is derived in 
ref.~\cite{Hetrick:1995wq}, 
\be
\label{anahetrick}
\frac{\Sigma^\mathrm{cont}}{e} = \frac{1}{4 \pi}\frac{(m_\pi^\mathrm{cont}/e)^2}{(m/e)}\; .
\ee

\vspace*{0.3cm}
\noindent {\bf Critical fermion mass}
\vspace*{0.3cm}

For Wilson fermions and for hypercube fermions, the  
fermion mass receives an
additive renormalization. Therefore, the  
critical values of the bare fermion mass, where e.g. the PCAC fermion 
mass vanishes has to be determined by non-perturbative simulations. 
We computed $m_\mathrm{PCAC}$, extracted from 
eq.~(\ref{pcacmass}) using the conserved currents of 
eqs.~(\ref{conservedVectorCurrent},\ref{conservedAxialCurrent})  
as a function of the bare
fermion masses, $m_0$ for Wilson and $m_\mathrm{hyp}$ for hypercube fermions.
We determined at each value of $\beta$ those values of $m_0$ and 
$m_\mathrm{hyp}$ where $m_\mathrm{PCAC}=0$. The value of these bare
fermion masses can be found in table~\ref{kappacrit}. Note that the 
actual 
critical mass values used for the twisted mass simulations 
in order to realize full twist, though compatible, differ slightly 
from the ones in table~\ref{kappacrit} since they were obtained with 
less statistics. 

\begin{table}
\begin{center}
\begin{tabular}{|l|l|l|}
\hline
$\beta$ & $-m_\mathrm{crit}^\mathrm{Wilson}$ &
          $-m_\mathrm{crit}^\mathrm{hyp}$ \\ \hline  
1.0	& 0.3204(7)	& 0.335(1) \\ \hline 
2.0	& 0.1968(9)	& 0.203(1) \\ \hline 
3.0	& 0.1351(2)	& 0.1392(4) \\ \hline 
4.0	& 0.1033(1)	& 0.1050(2) \\ \hline 
5.0	& 0.0840(1)	& 0.0856(1) \\ \hline 
6.0	& 0.0719(1)	& 0.0727(1) \\ \hline 

\end{tabular}
\caption{Critical hopping parameters at the values of $\beta$ used in our
simulations for the Wilson, $m_\mathrm{crit}^\mathrm{Wilson}$, and
hypercube, $m_\mathrm{crit}^\mathrm{hyp}$, fermions.}
\label{kappacrit}
\end{center}
\end{table}

\vspace*{0.3cm}
\noindent {\bf Fermion mass dependence of the pion mass}
\vspace*{0.3cm}

\begin{figure}[tbp]
\centering
\includegraphics[width=12.0cm]{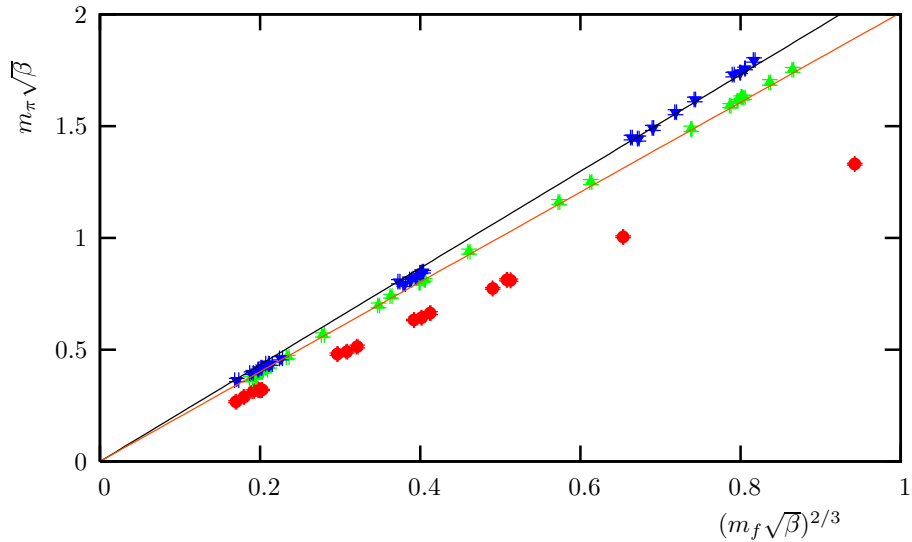}
\caption{Pseudo scalar mass $m_\pi\sqrt{\beta}$ versus 
$z=(m_f\sqrt{\beta})^{2/3}$ for three values of $\beta$. The solid lines are 
the analytical predictions of 
eqs.~(\ref{smilgaExpansion},\ref{gattrexpansion1}). Crosses represent our data
for $\beta=1$, triangles for $\beta=0.5$ and diamonds for $\beta=0.1$.}
\label{fig:mpimf}
\end{figure}

As described above, in the scaling analysis we will fix the scaling variable
$z=\left(m_f\sqrt{\beta}\right)^{2/3}$. To this end, we first explored the
dependence of the pion mass as a function of the PCAC fermion mass 
for fixed values of $\beta$. We give one example for the case of
hypercube fermions in fig.~\ref{fig:mpimf} using the conserved currents
of eqs.~(\ref{conservedVectorCurrent},\ref{conservedAxialCurrent})  
to compute the PCAC fermion mass. We also plot in the graph the analytic 
predictions of eqs.~(\ref{smilgaExpansion},\ref{gattrexpansion1}) 
for the fermion mass dependence of the pion mass.

The data in fig.~\ref{fig:mpimf} are for $\beta=0.1$, $\beta=0.5$ and 
$\beta=1.0$. Clearly, for $\beta=0.1$ the data do not follow the 
theoretical expectation while for the other values of $\beta$ there seems to
be some agreement with the analytical formulae. However, in order to make
more definite statements, a closer look to the scaling behaviour is 
clearly needed. Anyhow, the main purpose of fig.~\ref{fig:mpimf} is to show
that we have collected data that are concentrated around the values
of $z=0.2$, $z=0.4$ and $z=0.8$ that we are interested in for 
our final scaling
analysis. When the values of $z$ did not coincide directly 
with the desired ones, we 
performed an only very small linear interpolation of our data to achieve the 
exact values of $z$.

\vspace*{0.3cm}
\noindent {\bf Finite size effects}
\vspace*{0.3cm}

One source of a systematic error in the determination of $m_\pi$ are possible
finite size effects. For two dimensions the asymptotic finite size 
corrections for the
pseudo scalar mass were computed in ref.~\cite{Luscher:1985dn} and studied
numerically 
in the case of the Schwinger model in 
ref.~\cite{Gutsfeld:1999pu}. 
A very good agreement to the theoretical prediction
was found and it was observed that the variable $m_\pi L$ needs to be 
surprisingly large to suppress finite size corrections. 

\begin{figure}[tbp]
\centering
\includegraphics[width=12.0cm]{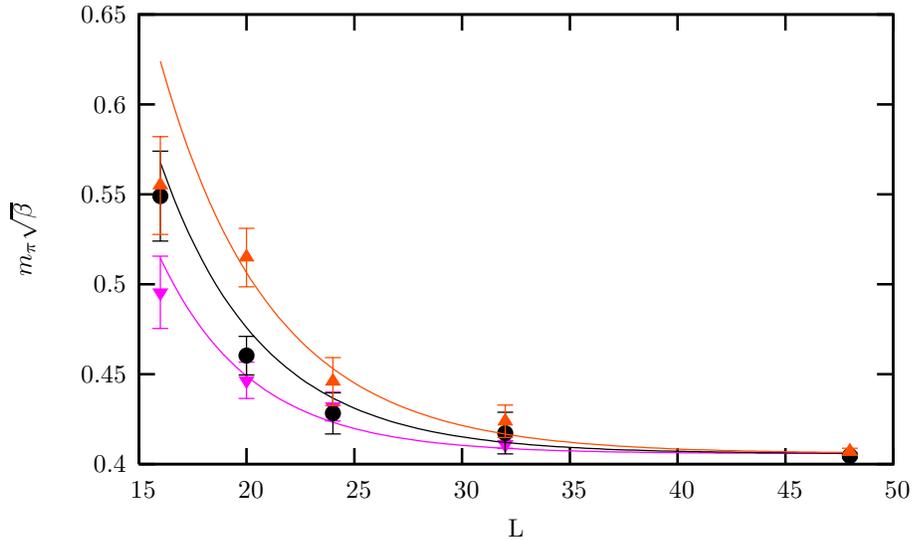}
\caption{We show $m_\pi\sqrt{\beta}$ as a function of $L$ at the example 
of Wilson fermions and three values of $\beta$ (downward triangles: 
$\beta=4$, circles: $\beta=5$, upward triangles: $\beta=6$). 
The fit curves, represented 
by solid lines, are obtained by performing a common fit to 
eq.~(\ref{fse}) to our 
data for $z=0.2$ and $z=0.4$ and 
for Wilson, hypercube and twisted mass fermions, choosing always
$\beta\ge 4$.}  
\label{fig:fse}
\end{figure}

Taking this as a warning, we followed the procedure of 
ref.~\cite{Gutsfeld:1999pu} and performed a finite size correction of our
values for the pseudo scalar masses when necessary.
We used the analytical formula 
\be
m_\pi(L)=m_\pi^\infty+A\frac{\sqrt{m_\pi^\infty}}{\sqrt{L}}e^{-Lm_\pi^\infty}
\label{fse}
\ee
where $m_\pi^\infty$ denotes the pseudo scalar mass in the infinite volume
limit. 
Rescaling eq.~(\ref{fse}) by $\sqrt{\beta}$ this formula can be applied 
to all our finite volume data at various, large enough values of 
$\beta$ to keep the effects of the lattice spacing small. 
Inspecting our data, we decided 
to perform
a global fit to our simulation data obtained with 
Wilson, hypercube and Wilson twisted mass fermions at $z=0.2$ and $z=0.4$ for
$\beta\ge 4$. As a result we found a 
``universal'' constant $A=12.4(6)$ which is compatible with the value computed
in ref.~\cite{Gutsfeld:1999pu}. 
In fig.~\ref{fig:fse} we give an example for the resulting fit for 
a subset of our data in the case
of Wilson fermions. 

We then examined all pseudo scalar masses that we will use in the detailed
scaling analysis described below and used eq.~(\ref{fse}) to 
analytically correct for finite size effects. We give in  
tables~\ref{wilsonresults}, \ref{hypercuberesults} and
\ref{tmwilsonresults}
the infinite volume values 
for the pseudo scalar mass
as obtained from this 
finite size correction. Note that in most of the cases, this correction 
is negligibly small and stays anyhow on the percent level.

\begin{table}
\begin{center}
\begin{tabular}{|l|l|l|l|l|l|l|l|}
\hline \multicolumn{8}{|l|}{Wilson fermion results} \\ \hline
$z$  &  $\beta$  &  $m_0^\mathrm{local}$  &  $m_0^\mathrm{cons}$  &  $L$  &  $m_{\pi,\infty}^\mathrm{local}\sqrt{\beta}$  &  $m_{\pi,\infty}^\mathrm{cons}\sqrt{\beta}$  &  $\Sigma_\mathrm{sub}^\mathrm{cons}\sqrt{\beta}$  \\ \hline
0.2 & 1.0 & -0.231367 & -0.230193 & 32 & 0.388(5) & 0.391(3) & 0.277(7) \\ \hline
0.2 & 2.0 & -0.132316 & -0.131842 & 32 & 0.402(3) & 0.406(2) & 0.252(6) \\ \hline
0.2 & 3.0 & -0.082626 & -0.081935 & 32 & 0.404(4) & 0.408(2) & 0.232(7) \\ \hline
0.2 & 4.0 & -0.057374 & -0.056979 & 32 & 0.406(2) & 0.408(2) & 0.231(8) \\ \hline
0.2 & 5.0 & -0.043199 & -0.043010 & 48 & 0.403(2) & 0.404(1) & 0.219(6) \\ \hline
0.2 & 6.0 & -0.034249 & -0.034009 & 48 & 0.404(2) & 0.406(2) & 0.225(4) \\ \hline
0.4 & 1.0 & -0.075308 & -0.059286 & 32 & 0.75(1) & 0.779(5) & 0.339(6) \\ \hline
0.4 & 2.0 & -0.017495 & -0.008352 & 32 & 0.783(8) & 0.813(6) & 0.339(3) \\ \hline
0.4 & 3.0 & 0.014505 & 0.019789 & 32 & 0.799(7) & 0.819(3) & 0.334(4) \\ \hline
0.4 & 4.0 & 0.026358 & 0.031078 & 32 & 0.80(1) & 0.820(3) & 0.340(3) \\ \hline
0.4 & 5.0 & 0.032526 & 0.035391 & 32 & 0.808(4) & 0.824(2) & 0.345(2) \\ \hline
0.4 & 6.0 & 0.036146 & 0.037693 & 32 & 0.811(3) & 0.820(2) & 0.345(2) \\ \hline
0.8 & 1.0 & 0.352443 & 0.537801 & 32 & 1.36(1) & 1.560(5) & 0.329(1) \\ \hline
0.8 & 2.0 & 0.314017 & 0.403487 & 32 & 1.52(1) & 1.690(5) & 0.400(1) \\ \hline
0.8 & 3.0 & 0.298677 & 0.350958 & 32 & 1.60(1) & 1.729(4) & 0.438(1) \\ \hline
0.8 & 4.0 & 0.276568 & 0.314133 & 32 & 1.63(1) & 1.745(5) & 0.469(1) \\ \hline
0.8 & 5.0 & 0.255009 & 0.285628 & 32 & 1.65(1) & 1.758(3) & 0.494(1) \\ \hline
0.8 & 6.0 & 0.239625 & 0.262974 & 32 & 1.686(6) & 1.770(3) & 0.518(1) \\ \hline
\end{tabular}

\caption{Results for the Wilson fermion simulations. The scaling
variable $\left(m_f\sqrt{\beta}\right)^{2/3}$ is fixed by using either the
local or the conserved currents to determine the PCAC fermion mass resulting
in the two different values of the pseudo scalar mass. 
Typical statistics of the runs were between $4000$ and $10000$ configurations.}
\label{wilsonresults}
\end{center}
\end{table}

\begin{table}
\begin{center}
\begin{tabular}{|l|l|l|l|l|l|l|l|}
\hline \multicolumn{8}{|l|}{Hypercube fermion results} \\ \hline
$z$  &  $\beta$  &  $m_0^\mathrm{local}$  &  $m_0^\mathrm{cons}$  &  $L$  &  $m_{\pi,\infty}^\mathrm{local}\sqrt{\beta}$  &  $m_{\pi,\infty}^\mathrm{cons}\sqrt{\beta}$  &  $\Sigma_\mathrm{sub}^\mathrm{cons}\sqrt{\beta}$  \\ \hline
0.2 & 1.0 & -0.237698 & -0.231044 & 32 & 0.39(1) & 0.416(3) & 0.290(6) \\ \hline
0.2 & 2.0 & -0.137038 & -0.133369 & 32 & 0.400(3) & 0.416(3) & 0.264(7) \\ \hline
0.2 & 3.0 & -0.085462 & -0.083199 & 32 & 0.400(7) & 0.415(2) & 0.250(6) \\ \hline
0.2 & 4.0 & -0.059364 & -0.058095 & 32 & 0.399(9) & 0.416(2) & 0.240(5) \\ \hline
0.2 & 5.0 & -0.044387 & -0.043885 & 48 & 0.408(2) & 0.410(2) & 0.246(8) \\ \hline
0.2 & 6.0 & -0.035528 & -0.034893 & 32 & 0.413(7) & 0.411(7) & 0.247(4) \\ \hline
0.4 & 1.0 & -0.083223 & -0.045098 & 32 & 0.78(2) & 0.844(9) & 0.375(2) \\ \hline
0.4 & 2.0 & -0.025253 & -0.004337 & 32 & 0.78(1) & 0.846(4) & 0.381(2) \\ \hline
0.4 & 3.0 & 0.012877 & 0.021147 & 32 & 0.806(6) & 0.836(3) & 0.382(2) \\ \hline
0.4 & 4.0 & 0.025787 & 0.031202 & 32 & 0.809(4) & 0.834(4) & 0.385(2) \\ \hline
0.4 & 5.0 & 0.031525 & 0.035539 & 32 & 0.812(3) & 0.831(2) & 0.392(3) \\ \hline
0.4 & 6.0 & 0.034082 & 0.037392 & 32 & 0.816(3) & 0.835(2) & 0.392(3) \\ \hline
0.8 & 1.0 & 0.311570 & 0.613589 & 32 & 1.39(2) & 1.742(6) & 0.366(1) \\ \hline
0.8 & 2.0 & 0.309020 & 0.433230 & 32 & 1.55(2) & 1.792(6) & 0.466(1) \\ \hline
0.8 & 3.0 & 0.293918 & 0.365237 & 32 & 1.61(1) & 1.793(4) & 0.517(1) \\ \hline
0.8 & 4.0 & 0.272609 & 0.321868 & 32 & 1.65(1) & 1.792(5) & 0.557(1) \\ \hline
0.8 & 5.0 & 0.255540 & 0.290832 & 32 & 1.672(7) & 1.792(4) & 0.588(1) \\ \hline
0.8 & 6.0 & 0.237752 & 0.266771 & 32 & 1.68(1) & 1.796(5) & 0.616(1) \\ \hline
\end{tabular}

\caption{Results for the hypercube fermion simulations. The scaling
variable $\left(m_f\sqrt{\beta}\right)^{2/3}$ is fixed by using either the
local or the conserved currents to determine the PCAC fermion mass resulting
in the two different values of the pseudo scalar mass. 
Typical statistics of the runs were between $4000$ and $10000$ configurations.}
\label{hypercuberesults}
\end{center}
\end{table}

\begin{table}
\begin{center}
\begin{tabular}{|l|l|l|l|l|l|l|l|}
\hline \multicolumn{8}{|l|}{Twisted mass fermion results} \\ \hline
$z$  &  $\beta$  &  $m_0^\mathrm{local}$  &  $m_0^\mathrm{cons}$  &  $L$  &  $m_{\pi,\infty}^\mathrm{local}\sqrt{\beta}$  &  $m_{\pi,\infty}^\mathrm{cons}\sqrt{\beta}$  &  $\Sigma_\mathrm{sub}^\mathrm{cons}\sqrt{\beta}$  \\ \hline
0.2 & 1.0 & -        & 0.071591 & 32 & -        & 0.434(2) & 0.303(7) \\ \hline
0.2 & 2.0 & 0.050499 & 0.054990 & 32 & 0.408(8) & 0.428(2) & 0.271(10) \\ \hline
0.2 & 3.0 & 0.045007 & 0.046886 & 32 & 0.411(2) & 0.422(2) & 0.249(6) \\ \hline
0.2 & 4.0 & -        & 0.041533 & 32 & -        & 0.421(1) & 0.226(6) \\ \hline
0.2 & 5.0 & 0.036965 & 0.037643 & 48 & 0.409(2) & 0.414(2) & 0.237(5) \\ \hline
0.2 & 6.0 & -        & 0.034651 & 48 & -       & 0.414(2) & 0.233(7) \\ \hline
0.4 & 1.0 & 0.177303 & 0.208213 & 32 & 0.78(1) & 0.874(3) & 0.446(7) \\ \hline
0.4 & 2.0 & 0.144653 & 0.157190 & 32 & 0.81(1) & 0.867(3) & 0.412(4) \\ \hline
0.4 & 3.0 & 0.125186 & 0.133453 & 32 & 0.819(8) & 0.851(2) & 0.394(4) \\ \hline
0.4 & 4.0 & 0.113557 & 0.117956 & 32 & 0.824(4) & 0.846(2) & 0.385(5) \\ \hline
0.4 & 5.0 & 0.102819 & 0.106827 & 32 & 0.820(4) & 0.844(2) & 0.388(4) \\ \hline
0.4 & 6.0 & 0.095373 & 0.098326 & 32 & 0.821(3) & 0.842(2) & 0.393(5) \\ \hline
0.8 & 1.0 & 0.471381 & 0.654028 & 32 & 1.47(2) & 1.799(5) & 0.571(2) \\ \hline
0.8 & 2.0 & 0.376485 & 0.459309 & 32 & 1.60(1) & 1.856(3) & 0.628(2) \\ \hline
0.8 & 3.0 & 0.334417 & 0.384597 & 32 & 1.66(2) & 1.838(3) & 0.638(2) \\ \hline
0.8 & 4.0 & 0.306509 & 0.337947 & 32 & 1.695(6) & 1.835(3) & 0.659(2) \\ \hline
0.8 & 5.0 & 0.282663 & 0.305270 & 32 & 1.719(6) & 1.829(3) & 0.671(2) \\ \hline
0.8 & 6.0 & 0.262653 & 0.280421 & 32 & 1.731(6) & 1.833(3) & 0.683(2) \\ \hline
\end{tabular}

\caption{Results for the Wilson maximally 
twisted mass fermion simulations. The scaling
variable $\left(m_f\sqrt{\beta}\right)^{2/3}$ is fixed by using either the
local or the conserved currents to determine the PCAC fermion mass resulting
in the two different values of the pseudo scalar mass. 
Typical statistics of the runs were between $4000$ and $10000$ configurations.}
\label{tmwilsonresults}
\end{center}
\end{table}

\begin{table}
\begin{center}
 \begin{tabular}{|l|l|l|l|l|l|l|}
   \hline
   \multicolumn{7}{|l|}{Overlap fermion results} \\ \hline
   $\beta$ &$m_\mathrm{ov}$& $(m_\mathrm{ov}\sqrt{\beta})^{2/3}$ &
   $(m_{f}^\mathrm{local}\sqrt{\beta})^{2/3}$ & $L$ &   
   $m_{\pi,\infty}^\mathrm{local}\sqrt{\beta}$ & $\Sigma\sqrt{\beta} $ \\
   \hline
   3.0 & 0.0464758 & 0.1864 & 0.194(15)& 24 & 0.37(4)   & 0.176(4) \\
   \hline
   4.0 & 0.0447214 & 0.2000 & 0.200(8) & 24 & 0.42(4)   & 0.177(4) \\
   \hline
   5.0 & 0.0420000 & 0.2066 & 0.20(3) & 28 & 0.46(4) & 0.138(2) \\
   \hline
   1.0 & 0.2529822 & 0.4000 & 0.4061(14) & 20 & 0.807(16) & 0.1950(16)\\
   \hline 
   2.0 & 0.1788854 & 0.4000 & 0.4069(19) & 20 & 0.824(14) & 0.2340(12) \\
   \hline
   3.0 & 0.1460595 & 0.4000 & 0.399(2) & 20 & 0.851(12) & 0.2518(7) \\
   \hline
   4.0 & 0.1264911 & 0.4000 & 0.383(2) & 20 & 0.833(10) & 0.2618(7)  \\
   \hline
   5.0 & 0.1142685 & 0.4027 & 0.396(5) & 24 & 0.837(10) & 0.2415(8)  \\
   \hline
   1.0 & 0.5724334 & 0.6894 & 0.797(2) & 20 & 1.433(10) & 0.2363(3)  \\
   \hline
   2.0 & 0.4232769 & 0.7103 & 0.8047(5) & 20 & 1.568(4) & 0.2905(3) \\
   \hline
   3.0 & 0.3553500 & 0.7236 & 0.7934(2) & 20 & 1.604(3) & 0.3270(3) \\
   \hline
   4.0 & 0.3152474 & 0.7353 & 0.7916(1) & 20 & 1.638(2) & 0.3549(3) \\
   \hline
   5.0 & 0.2916839 & 0.7521 & 0.8010(2) & 20 & 1.682(2) & 0.3811(6) \\
   \hline
 \end{tabular}

\caption{Results for the overlap fermion simulations. The scaling
variable $\left(m_f\sqrt{\beta}\right)^{2/3}$ is fixed by using the
local currents to determine the PCAC fermion mass.          
We give for comparison also the values of the scaling variable $z$ 
when the bare overlap fermion mass $m_\mathrm{ov}$ is used. 
Typical statistics of the runs were $4000$ configurations.}
\label{overlapresults}
\end{center}
\end{table}

\subsection{Results for the pseudo scalar mass}

\begin{figure}[tbp]
\centering
\includegraphics[width=12.0cm]{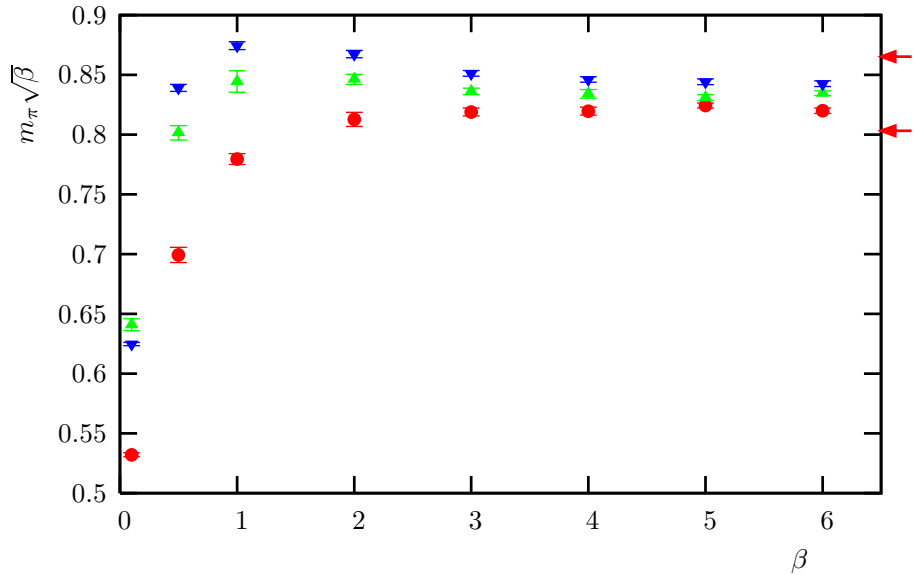}
\caption{We show $m_\pi\sqrt{\beta}$ as a function of $\beta$ at fixed
value of $z=0.4$ as determined from the PCAC fermion mass  
employing conserved currents. We represent results from Wilson fermions 
by circles, from twisted mass fermions by downward and for hypercube fermions
by upward triangles. The arrows represent the theoretical prediction, 
the upper arrow is from eq.~(\ref{gattrexpansion1}) the lower from 
eq.~(\ref{smilgaExpansion}).}
\label{fig:allbeta}
\end{figure}

In tables~\ref{wilsonresults}, \ref{hypercuberesults} and 
\ref{tmwilsonresults}
we give our results for the pseudo scalar masses at fixed scaling variable
$z$ of eq.~(\ref{z}), where $z$ is fixed either by using local or conserved 
currents to extract the PCAC fermion mass. While in the tables we give only
the results for $\beta\ge 1$, we have performed simulations also for smaller
values of $\beta$ for all actions, except overlap fermions. 
We show the result for $m_\pi\sqrt{\beta}$ as a function 
of $\beta$ in fig.~\ref{fig:allbeta}.
The graph is done for the example of $z=0.4$ where we used the conserved 
currents to determine the PCAC fermion mass.
Clearly, large lattice cut-off effects are observed when small values of 
$\beta < 1$ are taken. From the figure it is obvious that for an asymptotic
scaling analysis values of $\beta\ge 1$ should be taken. 

\begin{figure}[tbp]
\centering
\includegraphics[width=14.0cm]{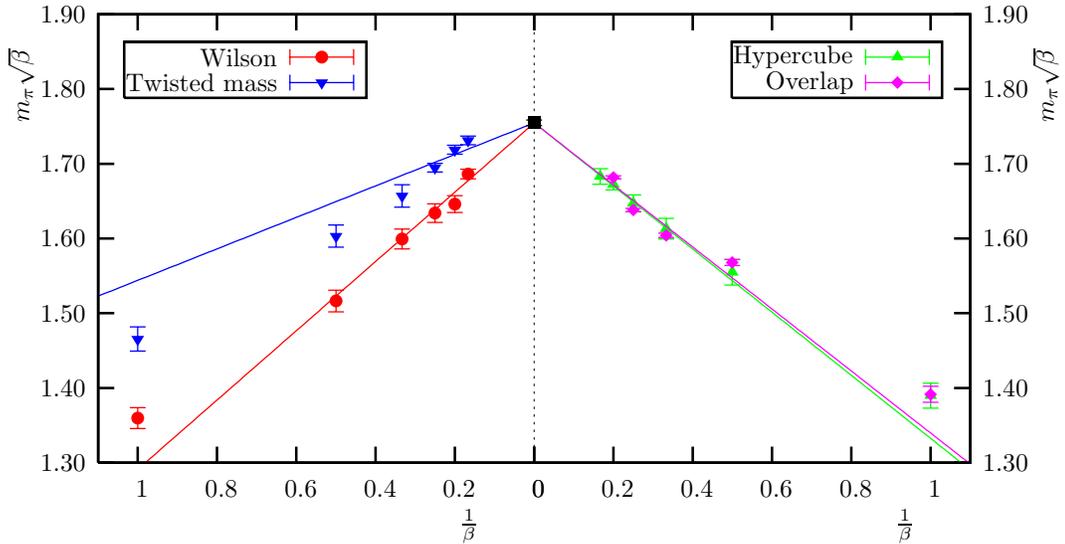}
\caption{Scaling test of $m_\pi\sqrt{\beta}$ as function of 
$1/\beta\propto a^2$ at fixed value of $z=0.8$. 
The solid lines 
represent linear constraint fits in $1/\beta$ demanding the same continuum limit
value for all kind of lattice fermions used.
For Wilson,
hypercube and twisted mass fermions we used data obtained at $\beta \geq 3$,
while for overlap fermions data with $\beta\geq 2$ were used.
}
\label{fig:scaling08}
\end{figure}

\begin{figure}[tbp]
\centering
\includegraphics[width=14.0cm]{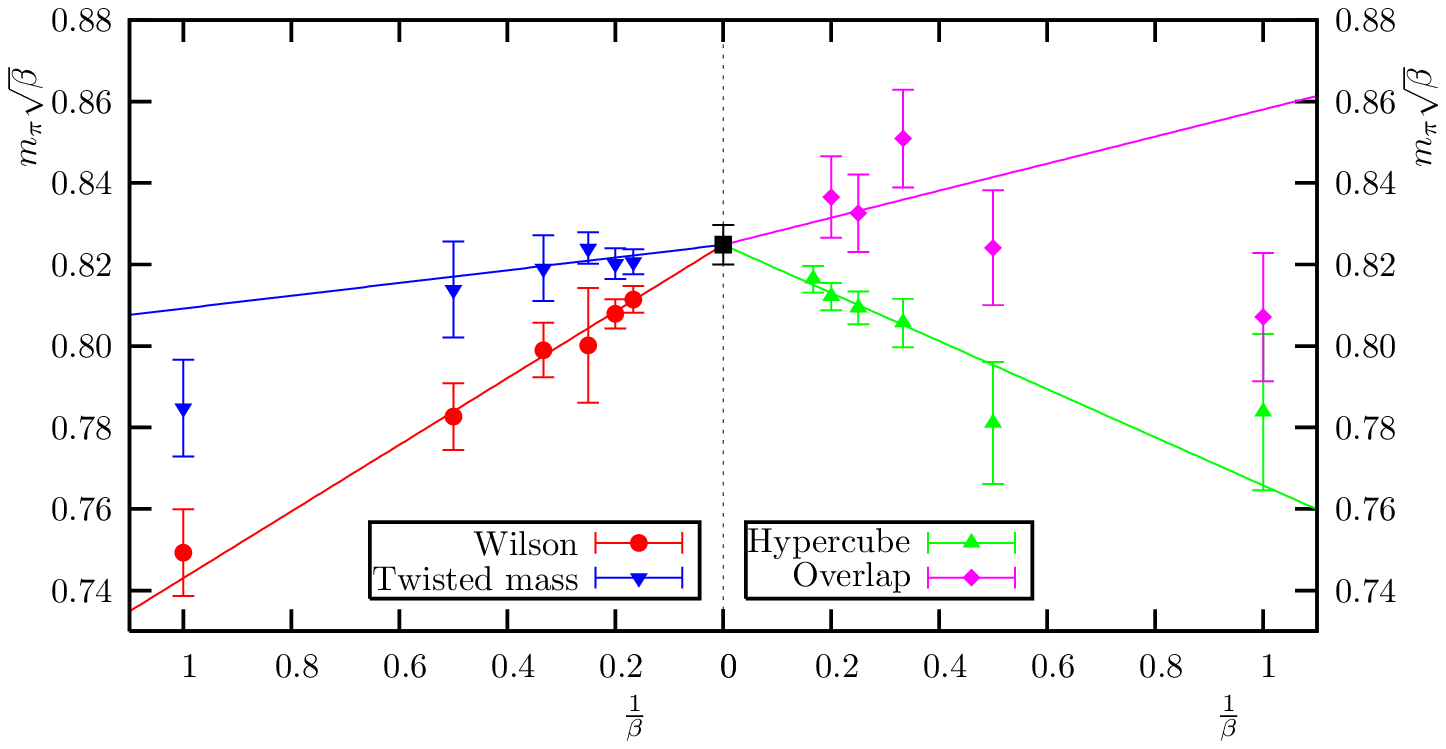}
\caption{Scaling test of $m_\pi\sqrt{\beta}$ as function of 
$1/\beta\propto a^2$ at fixed value of $z=0.4$. The solid lines are fits
explained 
in fig.~\ref{fig:scaling08}.} 
\label{fig:scaling04}
\end{figure}

\begin{figure}[tbp]
\centering
\includegraphics[width=14.0cm]{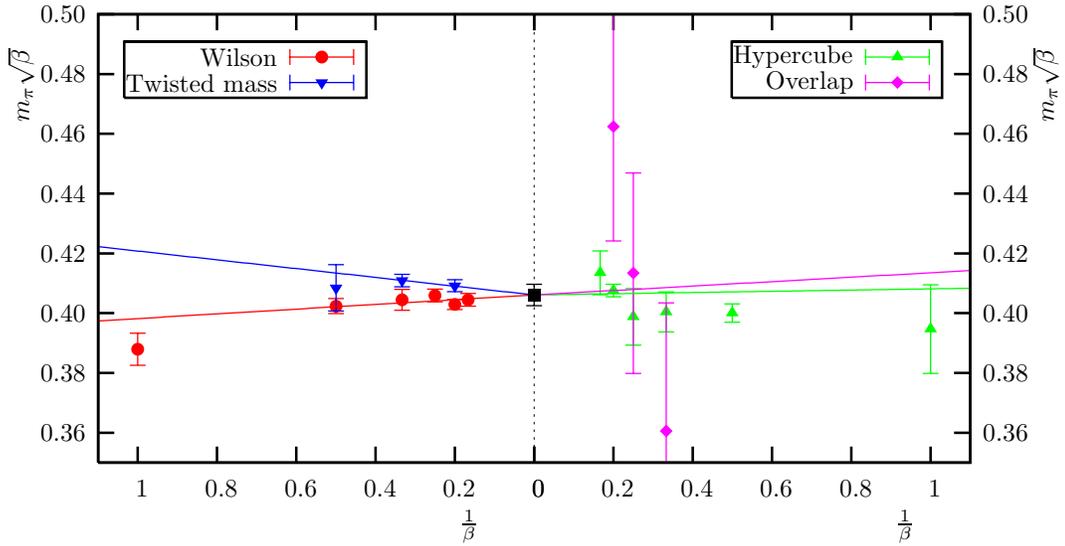}
\caption{Scaling test of $m_\pi\sqrt{\beta}$ as function of 
$1/\beta\propto a^2$ at fixed value of $z=0.2$. The solid lines are fits 
explained 
in fig.~\ref{fig:scaling08}.} 
\label{fig:scaling02}
\end{figure}

In figs.~\ref{fig:scaling08}, \ref{fig:scaling04} and \ref{fig:scaling02}
we show the results of our scaling test for fixed $z=0.8$, $z=0.4$ 
and $z=0.2$ respectively.
For the values of $\beta=1$ and $\beta=2$ at $z=0.2$, the determinant used
for reweighting the overlap results induced large fluctuations such that
no reliable values of physical observables could be extracted.
For all figures the value of $z$ was determined 
using local currents to compute the PCAC fermion mass. 
We show $m_\pi\sqrt{\beta}$ as a function of $1/\beta\propto a^2$. 
We first performed linear fits
in $1/\beta$ independently
for each lattice fermion used. 
Since these fits gave consistent continuum values for 
$m_\pi\sqrt{\beta}$ for all actions, see the example for $z=0.4$ in 
ref.~\cite{Christian:2005gd}, we finally performed constraint fits, 
demanding that all actions give the same continuum value
for $m_\pi\sqrt{\beta}$ at fixed value of $z$. We show these constraint fits 
in figs.~\ref{fig:scaling08}, \ref{fig:scaling04} and \ref{fig:scaling02}
as the solid lines. 
The data for all kind of fermions, Wilson, hypercube, maximally 
twisted mass and 
overlap, are nicely consistent with a linear behaviour in $a^2$. 
While this is expected for maximally 
twisted mass and overlap fermions, this outcome 
is somewhat surprising for hypercube and Wilson fermions. Note that our
finding for Wilson fermions is, however, consistent with
the results in ref.~\cite{Hoelbling:1998zg}. 

\begin{figure}[tbp]
\centering
\includegraphics[width=12.0cm]{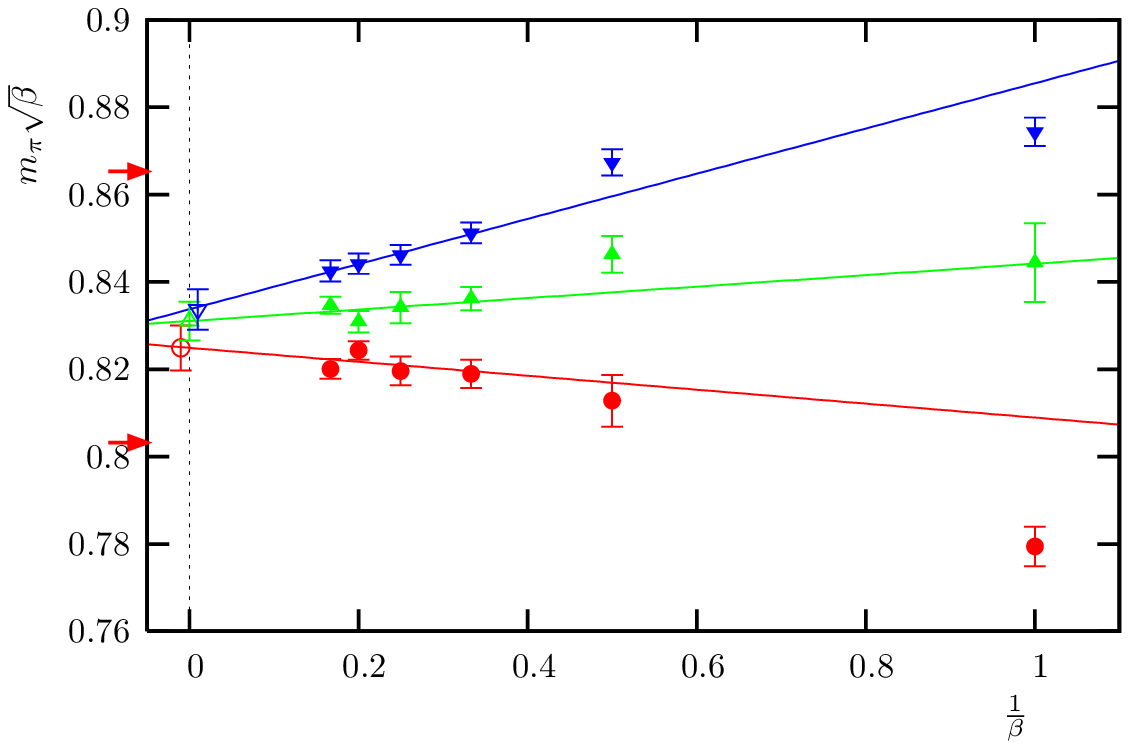}
\caption{Scaling of $m_\pi\sqrt{\beta}$ as function of $1/\beta$ 
using conserved currents to compute the PCAC fermion mass to fix 
$z=0.4$. We denote by circles Wilson, by downward triangles twisted mass 
and by upward triangles hypercube fermions. 
The solid lines 
represent linear fits in $1/\beta$ for $\beta \geq 3$.
The arrows represent the theoretical prediction,
the upper arrow is from eq.~(\ref{gattrexpansion1}) the lower from
eq.~(\ref{smilgaExpansion}).}
\label{fig:scalingcons}
\end{figure}

In general it is very difficult to decide which kind of lattice fermion
shows the best scaling behaviour and 
we cannot draw a definite conclusion 
here. First of all, all kind of fermions show only small lattice artefacts.
Second, if one would use the PCAC fermion mass from the conserved currents, 
the picture changes. We give an example for $z=0.4$ in 
fig.~\ref{fig:scalingcons} which reveals again an O(a$^2$) scaling 
for the here used Wilson, maximally twisted mass and hypercube fermions. 
When compared to fig.~\ref{fig:scaling04} where maximally twisted mass 
fermions show the smallest scaling violations, in fig.~\ref{fig:scalingcons}
hypercube fermions seem to do better. 

\vspace*{0.3cm}
\noindent {\bf Continuum comparison to theory}
\vspace*{0.3cm}

\begin{figure}[tbp]
\centering
\includegraphics[width=12.0cm]{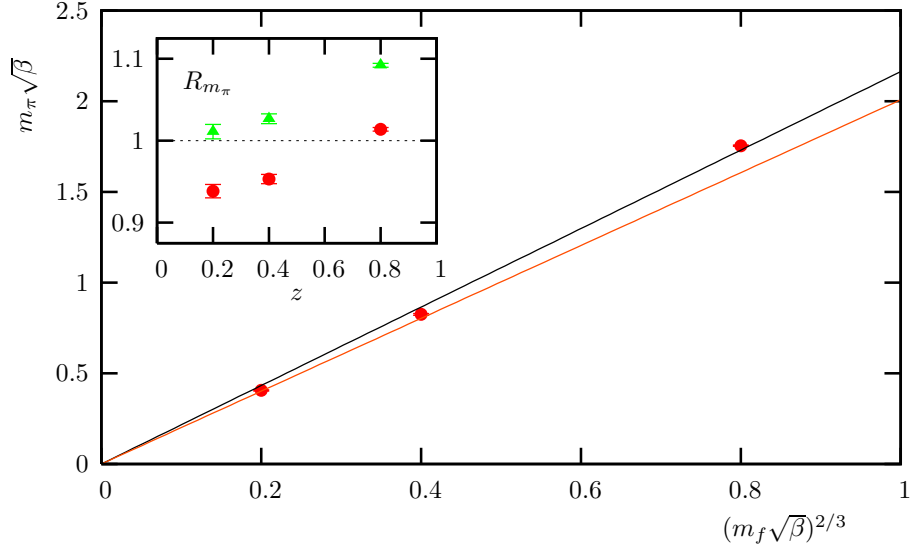}
\caption{Continuum behaviour of $m_\pi\sqrt{\beta}$ as function 
of z. The lower curve represents eq.~(\ref{smilgaExpansion}), the 
upper curve eq.~(\ref{gattrexpansion1}). The inlay gives the ratio 
$R_{m_\pi}$ of our 
non-perturbatively obtained data and the two theoretical calculations. 
Triangles use eq.~(\ref{gattrexpansion1}) and circles use 
eq.~(\ref{smilgaExpansion}).} 
\label{fig:mpicont}
\end{figure}

As was shown above, all kind of fermions show a nice scaling behaviour 
that is linear in $a^2$ and give a universal continuum limit. 
In fig.~\ref{fig:mpicont} we show
$m_\pi\sqrt{\beta}$ as a function of $z$ {\em in the continuum} and compare
to the theoretical expectations of eq.~(\ref{smilgaExpansion}) (lower line)
and 
eq.~(\ref{gattrexpansion1}) (upper line). As an inlay we plot the ratio
$R_{m_\pi}=m_\pi^\mathrm{data}/m_\pi^\mathrm{theor}$ 
of our non-perturbatively obtained data extrapolated to the 
continuum and the two theoretical predictions. Triangles represent
the data divided by the corresponding value computed from 
eq.~(\ref{gattrexpansion1}) while circles represent the data 
divided by the corresponding value computed from
eq.~(\ref{smilgaExpansion}).
To the precision we could obtain in this work, 
the theoretical predictions do not
describe the non-perturbatively obtained simulation data satisfactory 
at all values of $z$.
Only for small values of the fermion mass ($z=0.2$) there seems to be 
some agreement with eq.~(\ref{smilgaExpansion}) while at $z=0.8$ 
eq.~(\ref{gattrexpansion1}) seems to hold only.

\subsection{Scalar Condensate}

Another physical quantity 
we will consider in this work is the scalar condensate 
$\Sigma\equiv \langle \bar\psi \psi\rangle$,
for which analytical predictions exist, see eq.~(\ref{anacond}) and 
eq.~(\ref{anahetrick}). 
A very simple way to 
calculate the scalar condensate is
to compute 
\be
\Sigma= \frac{1}{V} \sum_x \mathrm{Tr}D^{-1}(x,x)
\ee
using a stochastic method.
We denote the so computed values of $\Sigma$
as $\Sigma_\mathrm{direct}$. 
A severe drawback of this definition of $\Sigma$ is that, 
at least in the case of Wilson and hypercube fermions, 
from the mixing with the identity operator 
a divergent piece $\propto 1/a$ appears 
that needs to be subtracted {\it non-perturbatively}.

In the case of the twisted mass fermions at full twist, 
it is possible to use the operator 
$\langle \bar\psi \sigma_3\tau_3 \psi\rangle$, 
i.e. the 3rd component of the pseudo scalar operator 
(see eq.~(\ref{pseudoscalar})), 
for the calculation of the scalar condensate. 
This operator does not mix with the identity operator
and thus the divergent piece does not appear. 

Also for the overlap operator there
is a definition
$\Sigma_\mathrm{ov}\equiv \langle \bar\psi(1-\frac{a}{2}D_{\rm ov}^{(0)}\psi\rangle$
which subtracts of the divergent piece automatically. 
For overlap fermions, 
we calculated therefore 
directly the scalar condensate from the eigenvalues of $D_\mathrm{ov}$
\be
\Sigma_\mathrm{ov}= \frac{1}{V}\langle\sum_i \frac{1}{\lambda_i}
(1-\lambda_i(0)/2)\rangle \; ,
\label{sigmaov}
\ee
where $\lambda_i$ denotes an eigenvalue of the massive overlap operator 
and $\lambda_i(0)$ denotes an eigenvalue at zero fermion mass, i.e.
the eigenvalues of $D_{\rm ov}^{(0)}$.

A second way, which avoids the appearance of the divergent piece 
from the beginning,
is to compute a so-called subtracted scalar condensate 
$\Sigma_\mathrm{sub}$ using the integrated axial Ward-Takahashi identity 
\cite{Bochicchio:1985xa}, 
\be
\Sigma_\mathrm{sub} =
2m_{\mathrm{PCAC}}\sum_{x}\langle \mathcal{\tilde P}_x\mathcal{\tilde P}_0 \rangle \; . 
\label{sigmasub}
\ee
In the twisted mass case,
this corresponds to the integrated PCVC relation
\be
\Sigma_\mathrm{sub}=
\langle \bar\psi \sigma_3\tau_3 \psi \rangle_\mathrm{sub}=
2m_{\mathrm{PCVC}}\sum_{x}\langle\mathcal{\tilde P}_x\mathcal{\tilde P}_0 \rangle\; . 
\label{sigmasubtm}
\ee

\vspace*{0.3cm}
\noindent
{\bf Comparison of $\Sigma_\mathrm{direct}$ and $\Sigma_\mathrm{sub}$}
\vspace*{0.3cm}

\begin{figure}[h!tb] 
\centering
\includegraphics[width=11.0cm]{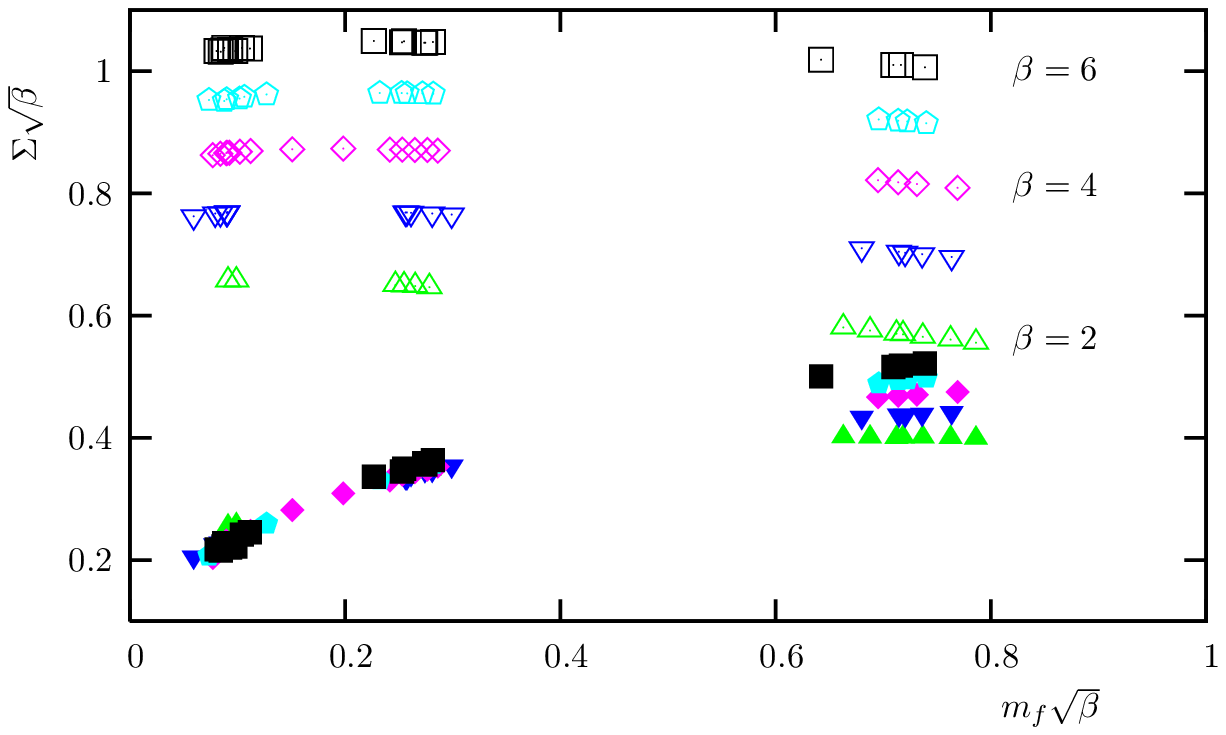}
\includegraphics[width=11.0cm]{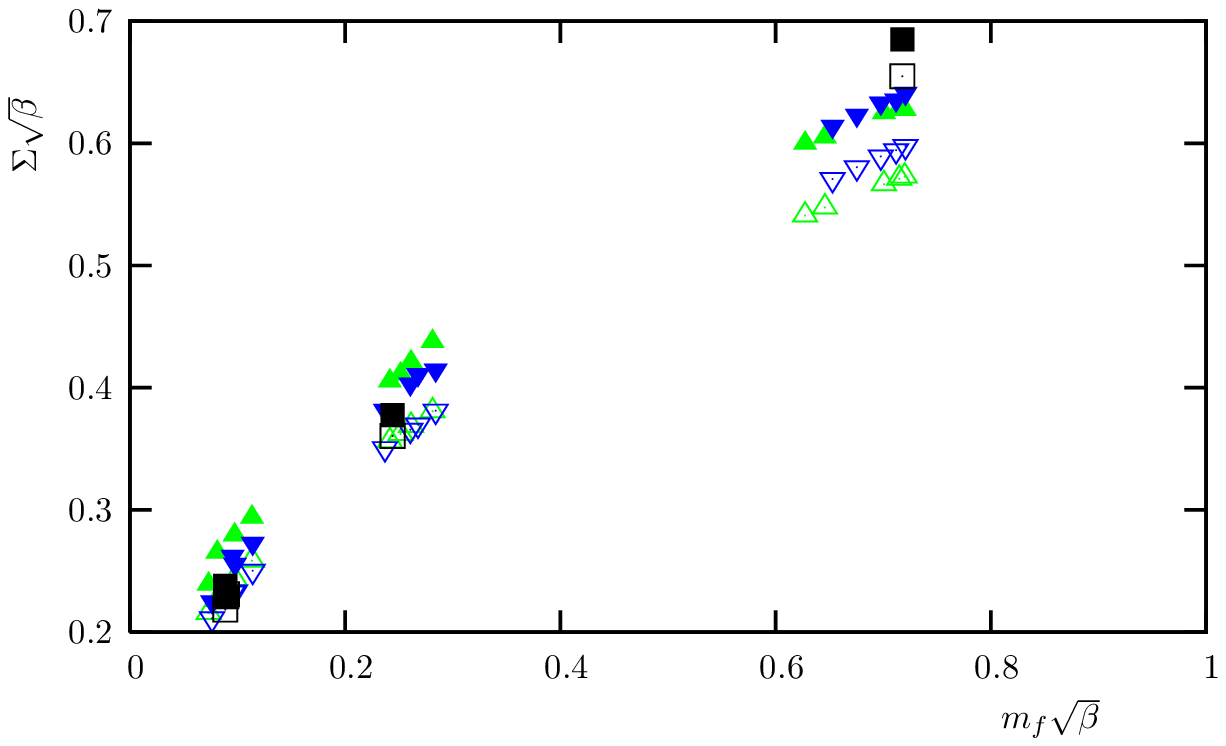}
\caption{
The condensate from the direct $\Sigma_\mathrm{direct}$ (open symbols) 
and the subtracted $\Sigma_\mathrm{sub}$ (full symbols) 
determinations. Error bars are within the size of the symbols.
Upper graph: $\Sigma_\mathrm{direct}$ and $\Sigma_\mathrm{sub}$
for Wilson fermion.
We indicate the values of $\beta$ where $\Sigma_\mathrm{direct}$ is obtained.
Lower graph: $\Sigma_\mathrm{direct}$ and $\Sigma_\mathrm{sub}$
for twisted mass fermions.
In this case, the results from both condensates are consistent up to
scaling violations.
}
\label{fig:directsub}
\end{figure}

To just illustrate the effect of 
using a direct and a subtracted definition 
of the scalar condensate, we plot 
$\Sigma_\mathrm{direct}$ and $\Sigma_\mathrm{sub}$ 
for Wilson and maximally twisted mass fermions 
in fig.~\ref{fig:directsub}.

In two dimensions, 
there is in general a relation between 
$\Sigma_\mathrm{direct}$ and $\Sigma_\mathrm{sub}$
given by 
\be
\Sigma_\mathrm{sub}=\Sigma_\mathrm{direct}-c_0/a \, .
\ee
The coefficient $c_0$ multiplying the $1/a$ divergence 
is to be determined non-pertur\-ba\-tively.
In the upper panel of fig.~\ref{fig:directsub},
the $1/a$ dependence is clearly visible for $\Sigma_\mathrm{direct}$, 
as for increasing values of $\beta$, the values of 
$\Sigma_\mathrm{direct}$ increase accordingly. 
It is clear from the figure that  
an extraction of a physical value for the scalar condensate will be 
very difficult since the term 
$c_0/a$ dominates the signal. 

According to \cite{Bochicchio:1985xa},
$c_0$ comes from the explicit breaking term of chiral symmetry
for Wilson fermions.
Therefore, one can expect 
that for twisted mass fermions at maximal twist 
this term is absent
and  
$\Sigma_\mathrm{direct}$ behaves like $\Sigma_\mathrm{sub}$.
This is shown in the lower panel of fig.~\ref{fig:directsub}.
In the case of twisted mass fermions both 
$\Sigma_\mathrm{direct}$ and $\Sigma_\mathrm{sub}$ are comparable 
(modulo scaling violations). 
In fact, we find a tendency that the ratio 
$\Sigma_\mathrm{sub}/\Sigma_\mathrm{direct}$
approaches one when $\beta$ is increased. 
We finally remark that also for the improved definition of the scalar 
condensate of eq.~(\ref{sigmaov})
in case of overlap fermions the divergence term is 
automatically subtracted of. 
As a result of this discussion we will 
calculate the scalar condensate from the subtracted scalar condensate 
in the case of Wilson, hypercube and maximally twisted mass 
fermions while for overlap fermions we will use
the direct calculation with the improved definition.




\vspace*{0.3cm}
\noindent{\bf Scaling of the scalar condensate}
\vspace*{0.3cm}

\begin{figure}[h!tb] 
\centering
\centerline{
\resizebox{7cm}{!}{\rotatebox{0}{\includegraphics{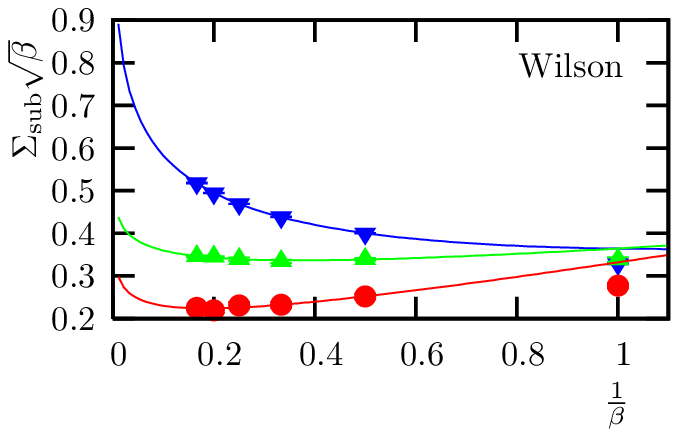}}}
\hspace{0.5cm}
\resizebox{7cm}{!}{\rotatebox{0}{\includegraphics{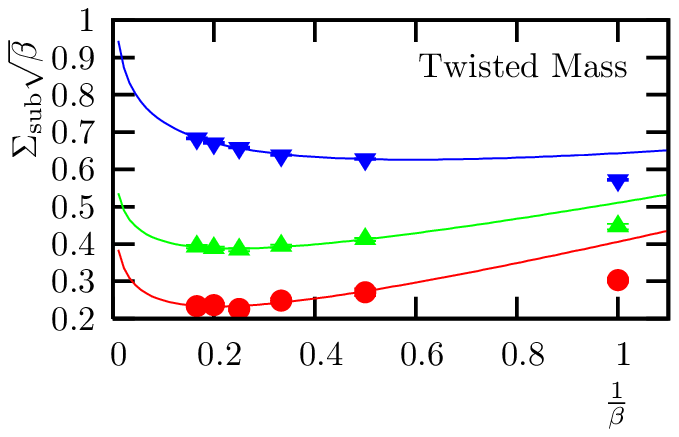}}}
}
\vspace{0.5cm}
\centerline{
\resizebox{7cm}{!}{\rotatebox{0}{\includegraphics{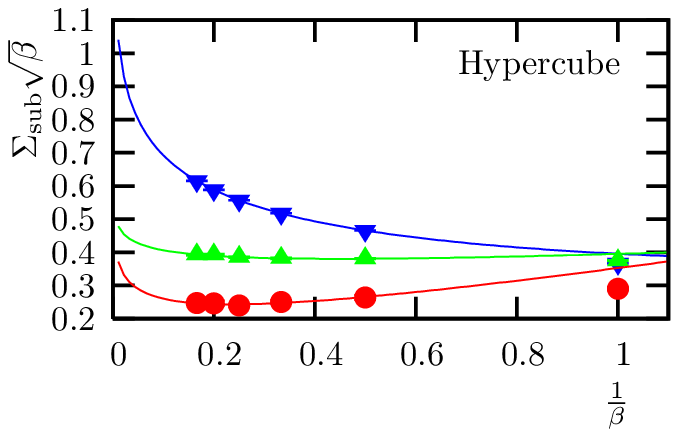}}}
\hspace{0.5cm}
\resizebox{7cm}{!}{\rotatebox{0}{\includegraphics{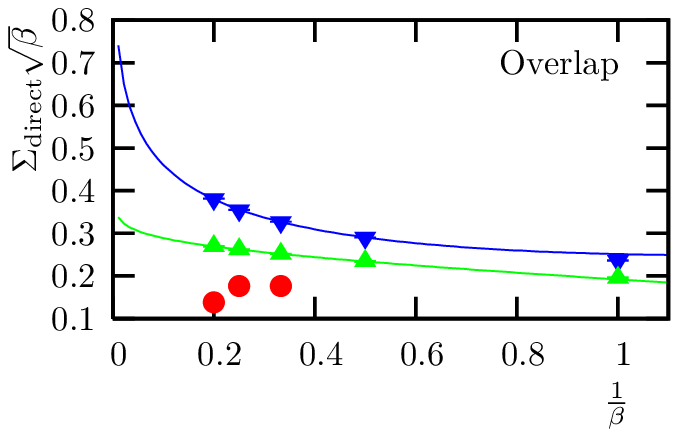}}}
}
\caption{
Scaling of the scalar condensate as a function of $1/\beta$ 
at $z=0.2$ (circles), 
$z=0.4$ (upper triangles) and $z=0.8$ (downward triangle). 
The solid lines represent the fits using eq.~(\ref{eq:scale-log}) for $\beta \geq 2$.
}
\label{fig:condfermion}
\end{figure}

Let us now turn to the results of the scaling behaviour of the scalar 
condensate
for the various fermion actions.
We show the 
results in fig.~\ref{fig:condfermion}.
$\Sigma_\mathrm{sub}$ is calculated for Wilson, hypercube and
twisted mass fermion,
and $\Sigma_\mathrm{direct}$ is done for overlap fermions,
at $z=(m_f \sqrt{\beta})^{2/3}=0.2$, 0.4 and 0.8.

The figures show a strong dependence of the scalar condensate on 
$1/\beta$, in particular for the heavy fermion region.
This behaviour can be explained by the fact that 
in two dimensions, the scalar condensate develops a logarithmic divergence
in $\beta$ when the lattice spacing is sent to zero  
\footnote{In 4 dimensions, there is also a logarithmic divergence.
But the origin is different and can be removed 
renormalising the scalar condensate by a multiplicative renormalisation factor.}.
This is most easily 
seen in the free theory 
where a simple computation of the scalar 
condensate leads to a behaviour (see also ref.~\cite{Durr:2004ta})
\be
\frac{\Sigma_\mathrm{free}}{e}
\propto \frac{m}{e} \log\left(\frac{1/a^2+m^2}{m^2}\right)
= \frac{m}{e} \log\left(\frac{\beta+(m/e)^2}{(m/e)^2}\right)
\; ,
\label{sigmafree}
\ee
where we have inserted $\beta=1/a^2e^2$ and $m$ denotes the continuum 
fermion mass. 
Fixing $m/e$, as we do in this work, and approaching the continuum limit 
by letting $\beta\rightarrow\infty$, 
a logarithmic increase of the scalar condensate in $\beta$ will appear.  
Fig.~\ref{fig:condfermion} shows clearly such an behaviour for all actions
we have employed. 

It is therefore natural to use a fit ansatz 
of the form 
\begin{equation}
\Sigma=A+\frac{B}{\beta}+C\log{\frac{1}{\beta}} .
\label{eq:scale-log} 
\end{equation}

In fig.~\ref{fig:condfermion} we show also the fit to the data using 
eq.~(\ref{eq:scale-log}).
For $z=0.2$ we could not extract reliable values for $\Sigma$ at $\beta=1,2$
since the determinant induced very large fluctuations. Since we were then
left with only three data points for a three parameter fit, and we were
also not sure whether the values for $\Sigma$ are possibly affected by
finite size effects, we did not use the fit in eq.(47) for overlap
fermions at $z=0.2$.

As can be seen, this fit function provides a nice description of the 
numerical data. 
In principle, for the cases of twisted mass and overlap fermions the 
divergent piece can be subtracted of from the evaluation of this term in 
the free theory when the fermion mass is matched. 
This is, however, very difficult 
for the cases of Wilson and hypercube fermions 
since there the matching of the fermion mass is not unique. 
We give in table~\ref{tab_fit_log} the fit results using 
eq.~(\ref{eq:scale-log}).



\begin{table}
\begin{center}
\begin{tabular}{|c|c|c|c|c|}
\hline
  $\Sigma=A+\frac{B}{\beta}+C \log(\frac{1}{\beta})$ & & & & \\
\hline
\hline
 $(m \sqrt{\beta})^{2/3}=0.2$ & $A$ & $B$ &  $C$ & $\chi^2$/dof   \\
\hline
Wilson & 0.118(103) & 0.215(150) & -0.0397(447) & 0.56  \\
\hline
HYP    & 0.0938(898)& 0.259(134) & -0.0610(389) & 0.52  \\
\hline
TM     & 0.0352(1189)& 0.371(179)& -0.0767(518) & 1.7   \\
\hline
\hline
 $(m \sqrt{\beta})^{2/3}=0.4$ & $A$ & $B$ &  $C$ & $\chi^2$/dof   \\
\hline
Wilson & 0.258(52) & 0.107(76)  & -0.0396(229) & 0.67  \\
\hline
HYP    & 0.316(38) & 0.0790(543)& -0.0357(174) & 0.48  \\
\hline
TM     & 0.218(69) & 0.293(98)  & -0.0698(307) & 0.47  \\
\hline
OV     & 0.249(19) &-0.0565(271)& -0.0200(83)  & 0.13  \\
\hline
\hline
 $(m \sqrt{\beta})^{2/3}=0.8$ & $A$ & $B$ &  $C$ & $\chi^2$/dof   \\
\hline
Wilson & 0.235(14) & 0.129(21)  & -0.145(6) & 0.62  \\
\hline
HYP    & 0.311(14) & 0.0843(207)& -0.162(6) & 0.19  \\
\hline
TM     & 0.465(31) & 0.178(45)  & -0.106(14)& 2.4   \\
\hline
OV     & 0.128(10) & 0.136(18)  & -0.160(4) & 0.022  \\
\hline
\end{tabular}
\end{center}
\caption{Fit results for the scalar condensate using 
eq.~(\ref{eq:scale-log})}
\label{tab_fit_log}
\end{table}


\section{Conclusions}

In this paper we have tested four different lattice fermions 
in their approach to the continuum limit in the 2-dimensional massive
Schwinger model with $N_f=2$ flavours of dynamical fermions. At fixed
scaling variable $z= (m_f\sqrt{\beta})^{2/3}=0.2,0.4,0.8$ we have 
computed the pseudo scalar mass $m_\pi\sqrt{\beta}$ and the scalar condensate
$\Sigma$ for various values of $\beta=1/e^2a^2$. 

For all kind of fermions used, Wilson, hypercube, twisted 
mass and overlap fermions, the scaling behaviour of $m_\pi\sqrt{\beta}$ 
appears to be linear 
in $a^2$. While this is expected for twisted mass fermions at full twist as
realized here and overlap fermions, this result is somewhat surprising 
for Wilson and hypercube fermions where a linear dependence on the 
lattice spacing was expected. 
Of course, it might be that another quantity can show different lattice
artefacts and the scaling behaviour shows another dependence on the
lattice spacing.

In fig.~\ref{fig:mpicont} we show our final results for $m_\pi\sqrt{\beta}$
computed non-perturbatively 
and extrapolated to the continuum such that a direct comparison 
to analytical predictions can be made. Only for a value 
of $z=0.2$ there seems to be a consistency with  
eq.~(\ref{smilgaExpansion}) valid for  strong couplings and small masses
while at $z=0.8$ 
eq.~(\ref{gattrexpansion1}), obtained from a large mass expansion, 
seems to describe the data.  
In general our conclusion is that to the precision we could compute our
results here, the analytical formulae do not describe the 
non-perturbatively obtained values of $m_\pi\sqrt{\beta}$
satisfactory at all values of $z$. 

As a second quantity we looked at the scalar condensate. We 
demonstrated that in our 2-dimensional setup
the use of a subtracted scalar condensate as derived 
from the integrated Ward identity is very useful to compute the scalar 
condensate since the so defined scalar condensate is free of 
divergence terms $\propto 1/a$. 
Our data are also consistent 
with  a logarithmic divergence in $a$ as can be derived in the free theory.

We also discussed some attempts to simulate the overlap operator dynamically
by using an infrared safe kernel to construct an 
approximate overlap operator. This approximate overlap 
operator is then used in the simulation and
physical observables are corrected by reweighting with 
the determinant ratio of the exact to the approximate operator.
We tested
this idea by computing stochastically this determinant ratio. For our best 
candidate for an approximate overlap operator in eq.~(\ref{signmod}),
for which we use an explicit infrared regulator in the sign function,
we found that the eigenvalue spectra of the exact and approximate operators
are very similar, see fig.~\ref{fig:lemons}, which shows 
the lemon shaped difference
spectra for various accuracies of the approximation corresponding to 
different values of $\delta$ in eq.~(\ref{signmod}). 
Nevertheless, even in this case, the fluctuations in the determinant ratio 
appeared to be very large even for a small value of the parameter 
$\delta$. 

Although this led us to conclude that this stochastic way of incorporating the
correction of the determinant ratio is not successful, we are 
still exploring to use
the approximate overlap operator as the guidance Hamiltonian 
in the molecular dynamics part of the 
Hybrid Monte Carlo simulation while for the accept/reject Hamiltonians 
the exact overlap operator is used.  
We tested this setup 
employing the hypercube operator as an approximate overlap operator 
as the guidance Hamiltonian in 
the molecular dynamics part. We found that even with this operator, which 
appeared to be the worst choice in the case of the stochastic estimate
of the determinant ratio tested here, 
the acceptance rates look reasonable, see ref.~\cite{Christian:2005gd}.  
We are investigating this promising result further at the moment with 
the other approximate overlap operators discussed in this paper.

\section{Acknowledgments}
We thank W. Bietenholz, V. Linke, C.~Urbach, U.~Wenger 
for many useful discussions. 
In particular we thank C.B. Lang for pointing out to us to use the
modified sign function of eq.~(\ref{signmod}).
The computer centers at DESY, Zeuthen,  and at the Freie Universit\"at 
in Berlin supplied us with 
the necessary technical help and computer
resources. 

\bibliographystyle{h-physrev4}
\bibliography{schwinger}

\end{document}